\documentclass[review,12pt]{elsarticle}

\usepackage{setspace}
\usepackage{subcaption}
\usepackage{multirow}
\usepackage{multicol}
\usepackage{graphicx}
\RequirePackage{amsmath}
\usepackage{bm}
\usepackage{xfrac}
\usepackage{float}
\usepackage{arydshln} %for dashed lines in tables

\usepackage{amsmath,esint,amsfonts}

\usepackage[most]{tcolorbox}
\usepackage[colorinlistoftodos]{todonotes}

\usepackage{lineno}
\usepackage[colorlinks=true,linkcolor=blue,urlcolor=blue,citecolor=blue,anchorcolor=blue]{hyperref}
\usepackage{cleveref} %for multiple figure references together
\journal{Journal}

\biboptions{sort&compress} %USE THIS
\begin{document}
\begin{frontmatter}

	\title{Sensitivity Analysis of Lift and Drag Coefficients for Flow over Elliptical Cylinders of Arbitrary Aspect Ratio and Angle of Attack using Neural Network}
	%Elsevier \LaTeX\ template\tnoteref{mytitlenote}}

	\author{Shantanu Shahane\fnref{Corresponding Author}$^{a,}$}
	\author{Purushotam Kumar$^{b}$}
	\author{Surya Pratap Vanka$^{c}$}
	\address{a National Center for Supercomputing Applications,\\
		University of Illinois at Urbana-Champaign, Urbana, Illinois 61801}
	\address{b Manufacturing Technology \& Engineering, \\Corning Incorporated, Painted Post, NY, 14870}
	\address{c Department of Mechanical Science and Engineering,\\
		University of Illinois at Urbana-Champaign, Urbana, Illinois 61801}
	\fntext[Corresponding Author]{\vspace{0.3cm}Corresponding Author Email Address: \url{shahaneshantanu@gmail.com}}
%	\fntext[UIUC]{Department of Mechanical Science and Engineering\\
%				University of Illinois at Urbana-Champaign, Urbana, Illinois 61801}
	%\author{Author Names}%\fnref{myfootnote}}



	\begin{abstract}
		Flow over bluff bodies has multiple engineering applications and thus, has been studied for decades. The lift and drag coefficients are practically important in the design of many components such as automobiles, aircrafts, buildings etc. These coefficients vary significantly with Reynolds number and geometric parameters of the bluff body. In this study, we have analyzed the sensitivity of lift and drag coefficients on single and tandem elliptic cylinders to cylinder aspect ratios, angles of attack, cylinder separation, and flow Reynolds number. Sensitivity analysis with Monte-Carlo algorithm requires several function evaluations, which is infeasible with high-fidelity computational simulations. We have therefore trained multilayer perceptron neural networks (MLPNN) using computational fluid dynamics data to estimate the lift and drag coefficients efficiently. Line plots of the variations in lift and drag as functions of the governing parameters are also presented. The present approach is applicable to study of various other bluff body configurations. \vspace{0.25cm}
	\end{abstract}


\end{frontmatter}

%\fulllengthtrue
%\fulllengthfalse

\section{Introduction}
The present study is concerned with the application of Machine Learning (ML) \cite{goodfellow2016deep,peng2020multiscale,schmidhuber2015deep} which seeks to explore physical phenomena through neural networks trained using data from experiments or numerical simulations. A typical study of machine learning consists of several initial numerical simulations or experiments with parameters chosen on a grid such as the Latin Hypercube \cite{mckay2000comparison,iman1981approach} spanning a multi-dimensional input space. The results of numerical simulations of the physical phenomena are then used to train a neural network with specified outputs. Once a neural network is trained and validated, it can then be used to quickly study characteristics of the phenomena in a multi-parameter space at low cost and optimize pre-determined objective functions. In comparison with actual full numerical simulations (or experiments), the neural networks execute much faster and can be efficiently used for sensitivity analysis, uncertainty quantification and multi-objective optimization.
\par In the recent years, there has been growing interest in the coupling of various architectures such as convolutional (CNN) \cite{han2019novel, miyanawala2017efficient, bukka2020assessment, ribeiro2020deepcfd, sekar2019fast, raissi2018hidden, ogoke2020graph, jin2018prediction,deng2019super} and multi-layer perceptron (MLPNN) \cite{sang2021data, seo2020direct, seo2021prediction, zhang2019numerical, alizadeh2020application, tang2020robust, shahane2020optimization, zhang2019quantifying, shahane2019uncertainty} neural networks with computational fluid dynamics and heat transfer. Here, we briefly review the publications which use MLPNN for surrogate modeling. \citet{sang2021data} used MLPNN to model drag on the rectangular cylinder for various aspect ratios, angles of attack and flow velocities. \citet{seo2020direct} trained a MLPNN to estimate the Nusselt number for three dimensional flow due to natural convection over sinusoidal cylinder. \citet{seo2021prediction} further modeled the Rayleigh-Benard natural convection induced by a circular cylinder placed insize a rectangular channel. \citet{zhang2019numerical} optimized the lift ot drag ratio on an airfoil using an upstream cylinder. \citet{alizadeh2020application} used a radial basis function MLPNN to investigate heat and mass transfer due to flow past a circular cylinder inside a porous media. \citet{tang2020robust} presented active control of flow over circular cylinder with deep reinforcement learning and MLPNN. \citet{shahane2020optimization} performed a multi-objective optimization of die cast parts combining genetic algorithm with MLPNN. \citet{zhang2019quantifying} developed physics informed neural networks to analyze uncertainty in direct and inverse stochastic problems. Further, \citet{shahane2019uncertainty} performed sensitivity and uncertainty analysis of three dimensional natural convection due to stochasticity in flow parameters and boundary conditions.

\par In this study, we apply the dense neural networks to study lift and drag due to flow over single and tandem elliptic cylinders at various angles of attack, aspect ratios and inter-cylinder spacing. Flows behind bluff bodies placed in a uniform fluid stream have been studied extensively in fluid mechanics literature. Numerous research papers exist on canonical geometries of a circular cylinder \cite{braza1986numerical, williamson1996vortex, dennis1970numerical, tritton1959experiments, zdravkovich1997flow, shahane2021high, shahane2021semi}, a square cylinder \cite{okajima1982strouhal,sharma2004heat,jackson1987finite,robichaux1999three,mukhopadhyay1992numerical}, and a flat plate \cite{najjar1995simulations,dennis1966steady,saha2007far}. Beginning with the early sketches by Da Vinci \cite{kemp2019leonardo}, the formation of a recirculation zone and the onset of unsteady shedding of vortices have been recognized as the distinct characteristics of flow behind an obstacle placed in a free stream. The alternate shedding of vortices from the shear layers due to the Kelvin-Helmholtz instability also gives rise to fluctuating lift and drag around the body caused by the alternating pressure forces around the body with a characteristic non-dimensional frequency called the Strouhal frequency. Numerous experimental, analytical, and numerical studies of flow over circular and rectangular cylinders have been previously reported. Computational studies ranging from very low Reynolds numbers to turbulent flows have been conducted for several canonical shapes using a variety of two-dimensional and three-dimensional numerical methods. Minimization of drag of automobiles and aircrafts while also considering stability issues has been extensively studied for propulsion systems.
\par Another canonical geometry that also possesses rich fluid physics and has practical relevance is an elliptical cylinder placed in a flowing fluid stream. The elliptical cylinder approaches the shapes of a flat plate for aspect ratio (ratio of maximum and minimum diameters, AR) of infinity, and a circular cylinder when AR is unity. Further, an elliptical cylinder can be placed at any arbitrary angular orientation with the free stream, providing another flow parameter in addition to aspect ratio and Reynolds number. Also, two or more elliptical cylinders can be placed in-line, or staggered, to passively control flow characteristics. Thus, elliptic cylinders provide a big parameter space to explore flow physics of fundamental and practical importance. Several studies have been conducted on wakes of elliptic cylinders but in comparison with those of circular and rectangular cylinders, such studies have been less in number.
\par \citet{shintani1983low} analyzed the low Reynolds number flow past an elliptical cylinder placed normal to a uniform stream using a method of matched asymptotic expansions. \citet{park1989flow} numerically studied the flow past an impulsively started slender elliptic cylinder (aspect ratio of 14.89) for a Reynolds number between 25 and 600. The stream function vorticity transport equations are solved in an elliptical boundary-fitted coordinate system. The angle of incidence is varied at small increments of $2.5^\text{o}$ between $0^\text{o}$ and $90^\text{o}$ at a fixed Reynolds number, and the flow regime is mapped. For small angles of attack, the flow is found to be steady up to Reynolds numbers of 300. \citet{raman2013effect} used a Cartesian grid code employing the immersed boundary method to compute wake of an elliptic cylinder with aspect ratio varying between 0.1 and 1.0 in steps of 0.1. The wake formation and transition to unsteady flow are presented as a function of aspect ratio for a fixed Reynolds number. \citet{paul2014numerical} further used the same numerical code and studied the effects of aspect ratio and angle of attack on wake characteristics of an elliptic cylinder. The different vortex shedding patterns are classified using the velocity and vorticity profiles. The lift, drag, and Strouhal number are also documented as a function of the cylinder aspect ratio, angle of attack, and the Reynolds number.
\par Several studies of flow over impulsively started elliptic cylinders are also reported. \citet{lugt1974laminar} computed the laminar flow over an abruptly accelerated cylinder at an angle of incidence of $45^\text{o}$. They observed steady flow only until Re = 30, and the von K\'arm\'an vortex shedding at Re = 200. \citet{taneda1972development} performed experiments of impulsively started elliptic cylinders and presented photographs of streak lines and streamlines. \citet{patel1981flow} performed a semi-analytic study of an impulsively started elliptic cylinder at various angles of attack and presented flow patterns downstream of the cylinder. \citet{ota1987flow} also conducted experiments of flow around an elliptic cylinder and presented the Reynolds number for formation of a separation zone and for onset of vortex shedding. \citet{nair1997unsteady} investigated high Reynolds number flow over an elliptical cylinder using a two-dimensional stream-function-vorticity formulation and third/fourth-order differencing of the convection terms. Effects of Reynolds number, angle of attack, and thickness to chord ratio are presented.
\par \citet{faruquee2007effects} studied the laminar flow over an elliptic cylinder in the steady regime using the commercial CFD code FLUENT for a Reynolds number of 40 based on the hydraulic diameter. They varied the aspect ratio from 0.3 to 1 with the major axis oriented parallel to the free stream. They reported that no vortices are formed below an aspect ratio of 0.34, after which a pair of symmetric vortices are formed. The wake size and drag coefficient are observed to increase quadratically with the aspect ratio at the fixed Reynolds number of 40. \citet{dennis2003steady} studied the steady flow over an elliptic cylinder with minor to major axes ratios of 0.2 and 0.1 and for Reynolds number between 1 and 40. Lift and drag coefficients and surface vorticity distributions are presented for different angles of inclination. A series truncation method described by \citet{badr2001numerical} is used. \citet{jackson1987finite} used a finite element method to study laminar flow past cylinders of various shapes including elliptic cylinders. In this study, emphasis is based on the Hopf bifurcation from a steady flow to an unsteady flow as the Reynolds number is increased. The transition Reynolds number is estimated to vary from 35.7 at zero angle of attack to 141.4 at $90^\text{o}$ for an ellipse of major to minor axis ratio of 2.0. \citet{d1995steady} studied the steady flow over an elliptic cylinder at different angles of inclination using a stream-function-vorticity approach on a curvilinear grid for Reynolds number of 5 and 20. This study is followed by \citet{d1999unsteady} in which the unsteady flow of an impulsively started elliptic cylinder translating about its axis is computed by the stream-function-vorticity approach. A similar study is also conducted by \citet{patel1981flow}.
\par In this study, we have first conducted a large set of numerical simulations of flow over single and tandem elliptic cylinders \cite{ota1986flow} placed in a uniform free stream. For a single cylinder, the aspect ratio defined as ratio of the major to minor axes is varied from 1 to 3, and the angle of inclination is varied counterclockwise from $0^\text{o}$ to $180^\text{o}$. For tandem inline cylinders, the aspect ratio and the angle of attack are varied independently for the two cylinders along with the separation. The ranges of aspect ratios, angles and separation are $[1,3]$, $[0^\text{o},180^\text{o}]$ and $[4,10]$ respectively. In addition, the free stream Reynolds number defined based on the free stream velocity and the major axis of the leading elliptical cylinder is varied in the range $[20,40]$. The Reynolds number is limited to the steady regime and a multi-layer perceptron neural network (MLPNN) is trained over a large data set spanning the three and six parameters for single and double cylinders respectively. The accuracy of the network is assessed on an unseen test data set. After ensuring acceptable accuracy, the network is applied to demonstrate the trends at different values of parameters and estimate sensitivity of lift and drag coefficients to input parameters. Since the neural network estimations run very fast, they can be performed interactively and can be used as virtual laboratories.
\par \Cref{Sec:Numerical Simulations of Flow over Elliptical Cylinders} briefly describes the commercial software COMSOL used to perform the initial CFD flow simulations. \Cref{Sec:MLPNN description} describes the architectures and error analyses of the developed neural networks. \Cref{Sec:Variation of Lift and Drag Coefficients} presents application of the neural network to analyze the observed trends in lift and drag coefficients. \Cref{Sec:Sensitivity Analysis} further uses the neural network to estimate their sensitivity to input parameters. A summary and future directions is given in \cref{Sec:Conclusion}.
\section{Numerical Simulations of Flow over Elliptical Cylinders}\label{Sec:Numerical Simulations of Flow over Elliptical Cylinders}
To perform flow computations, we have used the COMSOL software which solves the Navier-Stokes equations for the given set of flow variables. COMSOL is a finite element based solver for multi-physics simulations. We have used the steady state two-dimensional laminar incompressible flow (spf) module. It solves the continuity and momentum equations in the x and y directions. An unstructured mesh with triangular elements has been used to discretize the domain. Quadratic elements with second order accurate discretization schemes are chosen to represent the velocity and pressure variables. The direct solver option with full velocity-pressure coupling is selected to solve for the flow field.
\par The minor axis of the ellipse is taken as one unit in length and the length of the major axis is varied as per the aspect ratio selected for the desired case. The computational domain has a length of 90 units and a height of 24 units. The distance of the flow inlet to the center of the upstream ellipse is 24 units, and the second ellipse is placed downstream with the specified separation. The six variables namely Reynolds number, two major to minor axis ratios, two angles of attack of the ellipses and the separation between center of ellipses are chosen as input parameters. The upstream velocity is fixed at 1 m/s and the kinematic viscosity is changed according to the Reynolds number. Reynolds number is defined with respect to the major axis of the upstream ellipse.
\begin{figure}[H]
	\centering
	\begin{subfigure}[t]{0.36\textwidth}
		\includegraphics[width=\textwidth]{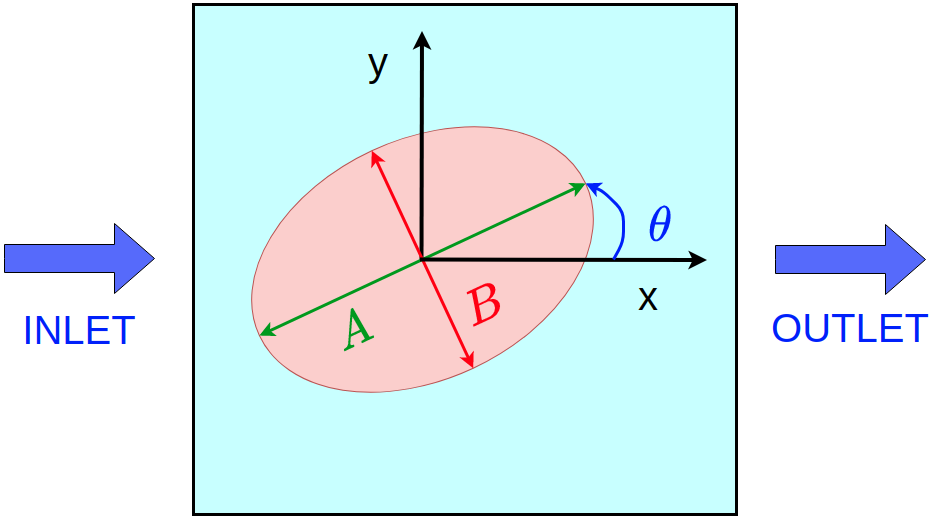}
		\caption{Case of Single Elliptic Cylinder}
		\label{Fig:Schematic of the Flow Domain single}
	\end{subfigure}
	\hspace{0.11\textwidth}
	\begin{subfigure}[t]{0.51\textwidth}
		\includegraphics[width=\textwidth]{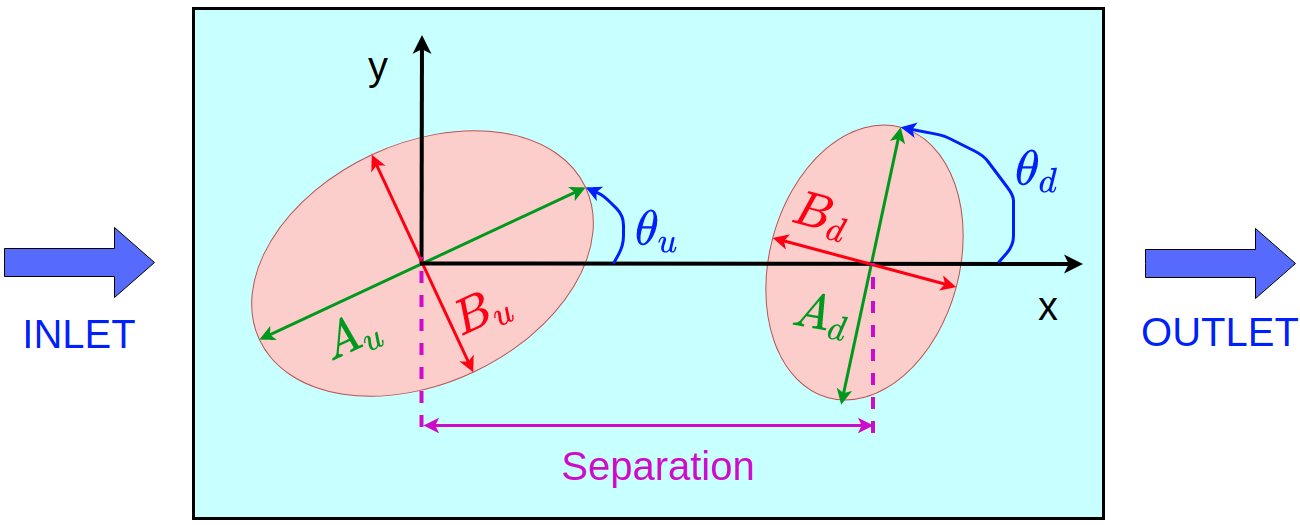}
		\caption{Case of Double Elliptic Cylinder}
		\label{Fig:Schematic of the Flow Domain double}
	\end{subfigure}
	\caption{Schematic of the Flow Domain in the Vicinity of the Cylinders (Cylinder Dimensions Artificially Enlarged)}
	\label{Fig:Schematic of the Flow Domain}
\end{figure}
\par The top and bottom boundaries are specified to have the free stream velocities with no-slip conditions on the surfaces of the elliptic cylinders. Uniform unit normal and zero transverse velocities are prescribed at the inlet. Constant pressure is imposed at the outlet. The drag and lift coefficients are defined as
\begin{equation}
	C_L = \frac{F_D}{0.5\rho U^2 A} \hspace{1cm} C_L = \frac{F_L}{0.5\rho U^2 A}
\end{equation}
where, $F_D$ and $F_L$ are drag and lift forces per unit length respectively and $A$ is taken to be the length of the major axis. Both pressure and viscous forces are included in calculation of drag and lift forces.
\par
\begin{figure}[H]
	\centering
	\begin{subfigure}[t]{0.32\textwidth}
		\includegraphics[width=\textwidth]{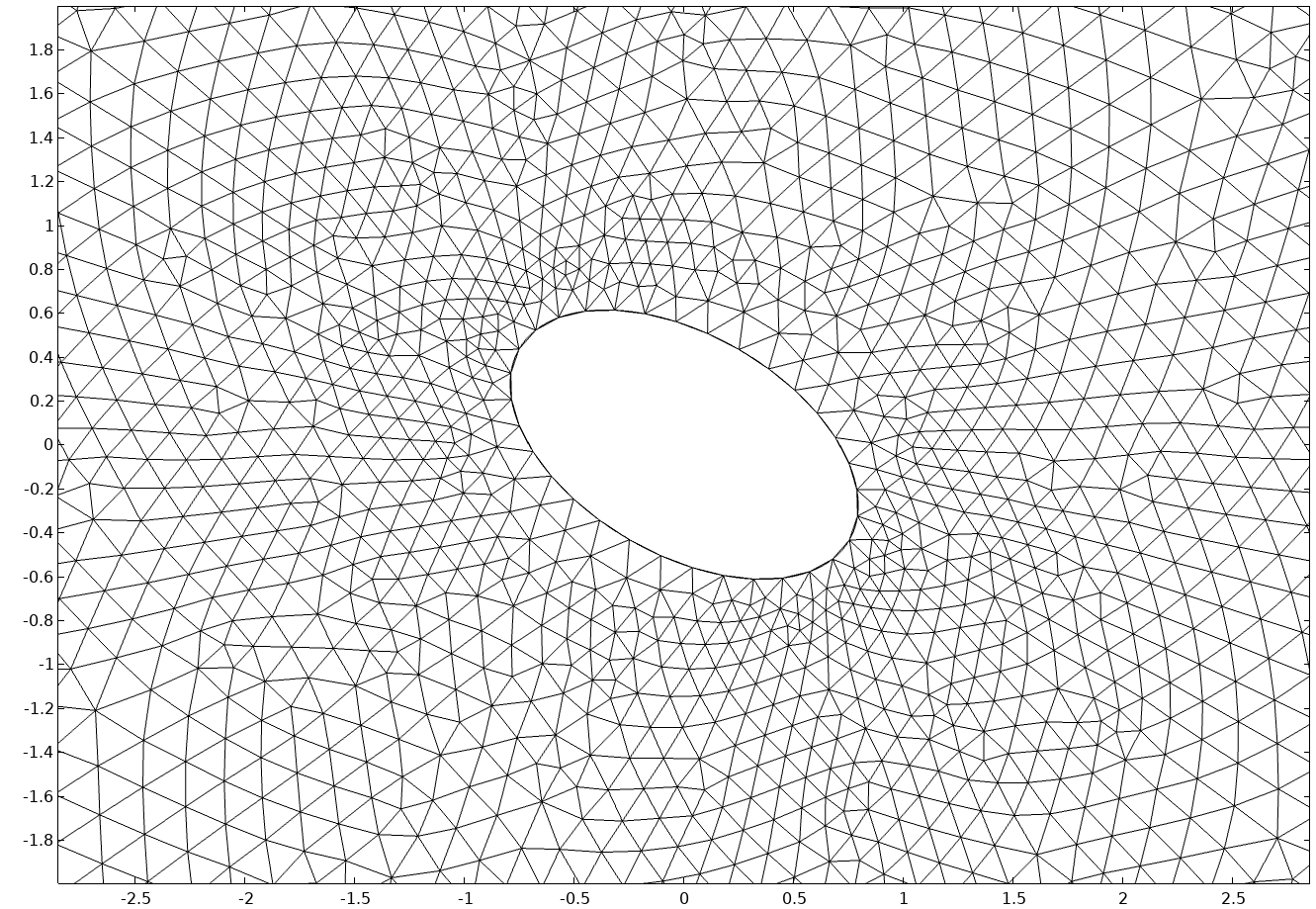}
		\caption{Coarse Mesh (32052 Elements)}
	\end{subfigure}
	\begin{subfigure}[t]{0.32\textwidth}
		\includegraphics[width=\textwidth]{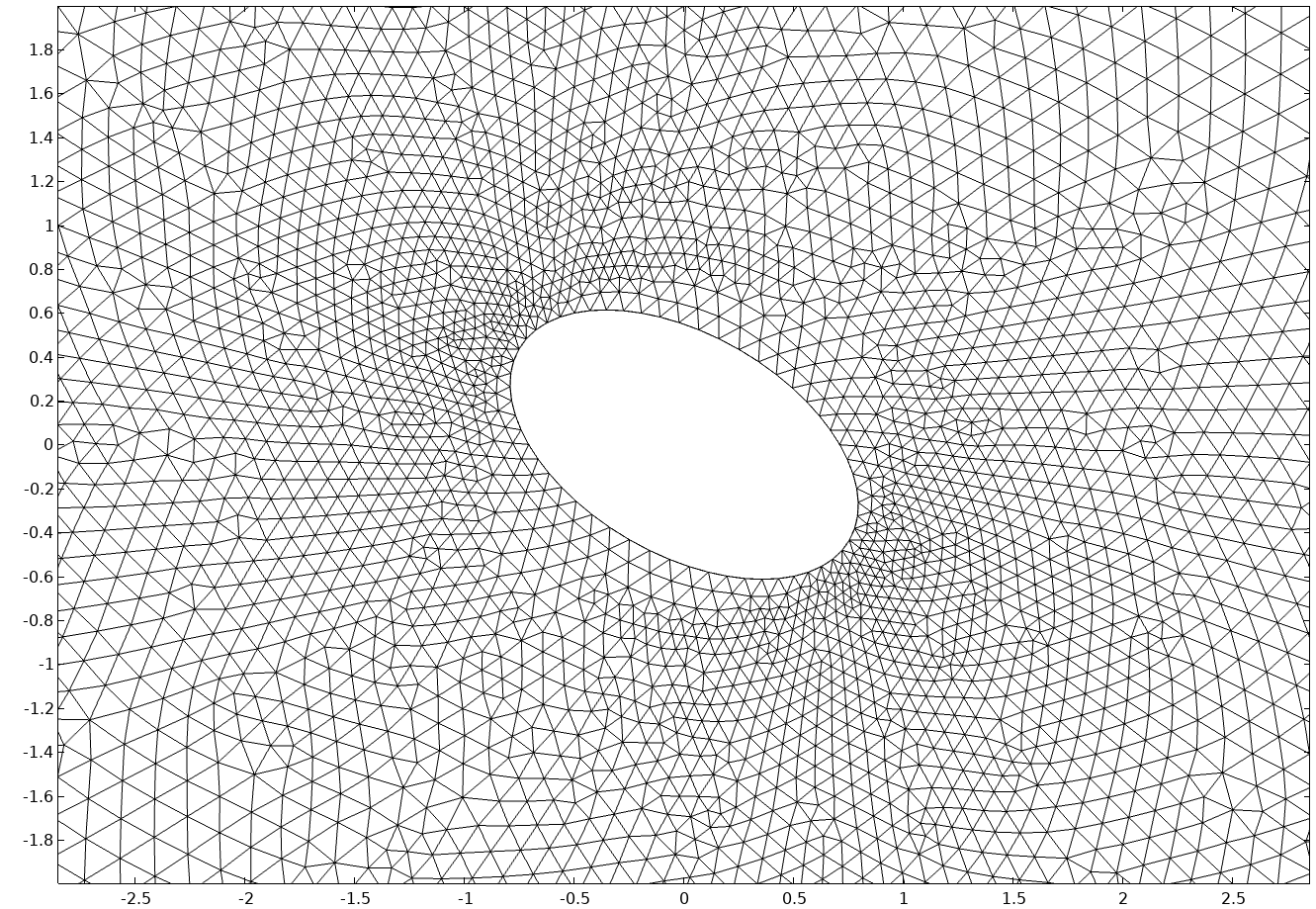}
		\caption{Medium Mesh (113198 Elements)}
	\end{subfigure}
	\begin{subfigure}[t]{0.32\textwidth}
		\includegraphics[width=\textwidth]{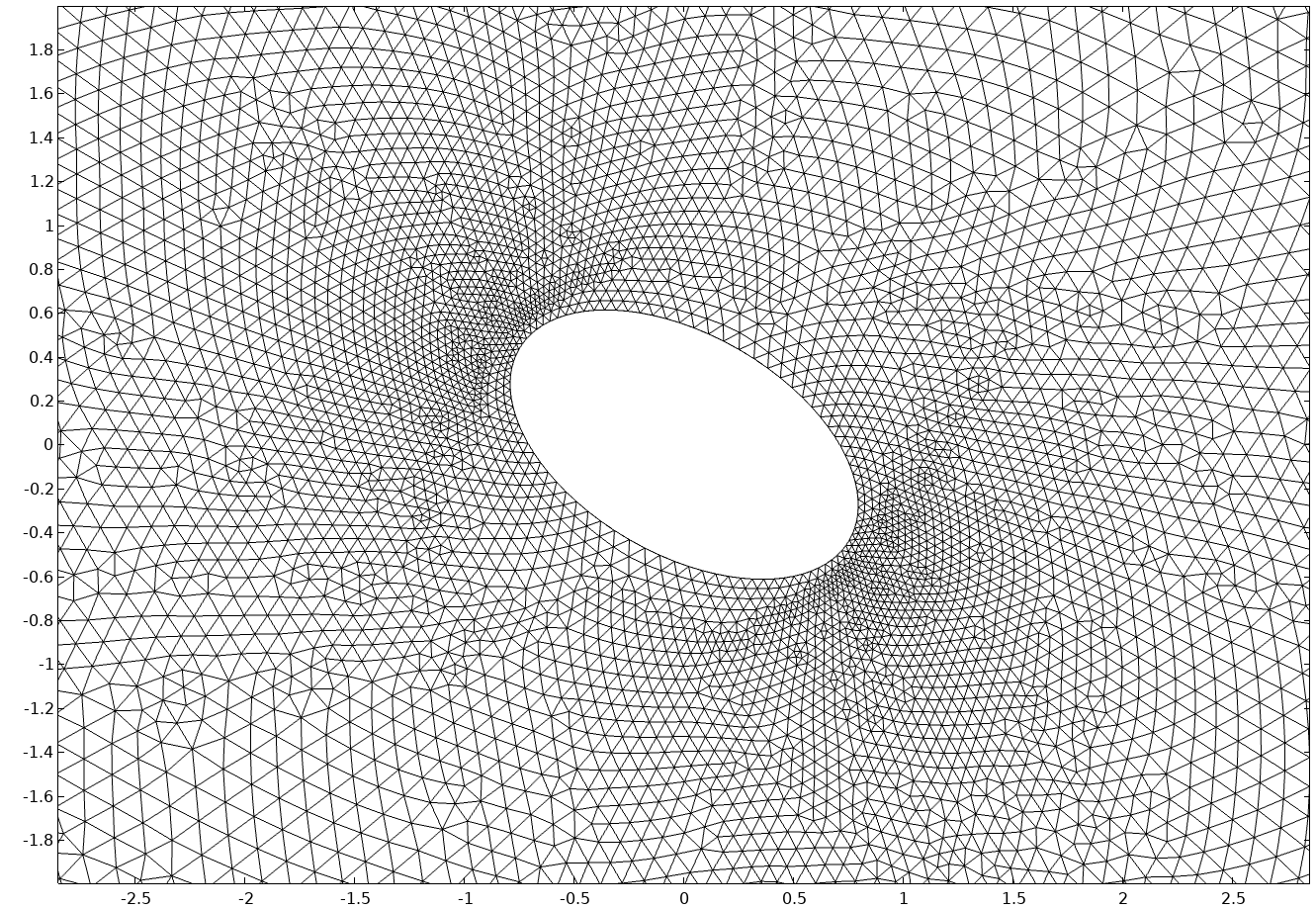}
		\caption{Fine Mesh (451098 Elements)}
	\end{subfigure}
	\caption{Single Cylinder Case: Successively Refined Meshes for Grid-Independence Study}
	\label{Fig: Single Cylinder Case: Grids}
\end{figure}

\begin{figure}[H]
	\centering
	\begin{subfigure}[t]{0.32\textwidth}
		\includegraphics[width=\textwidth]{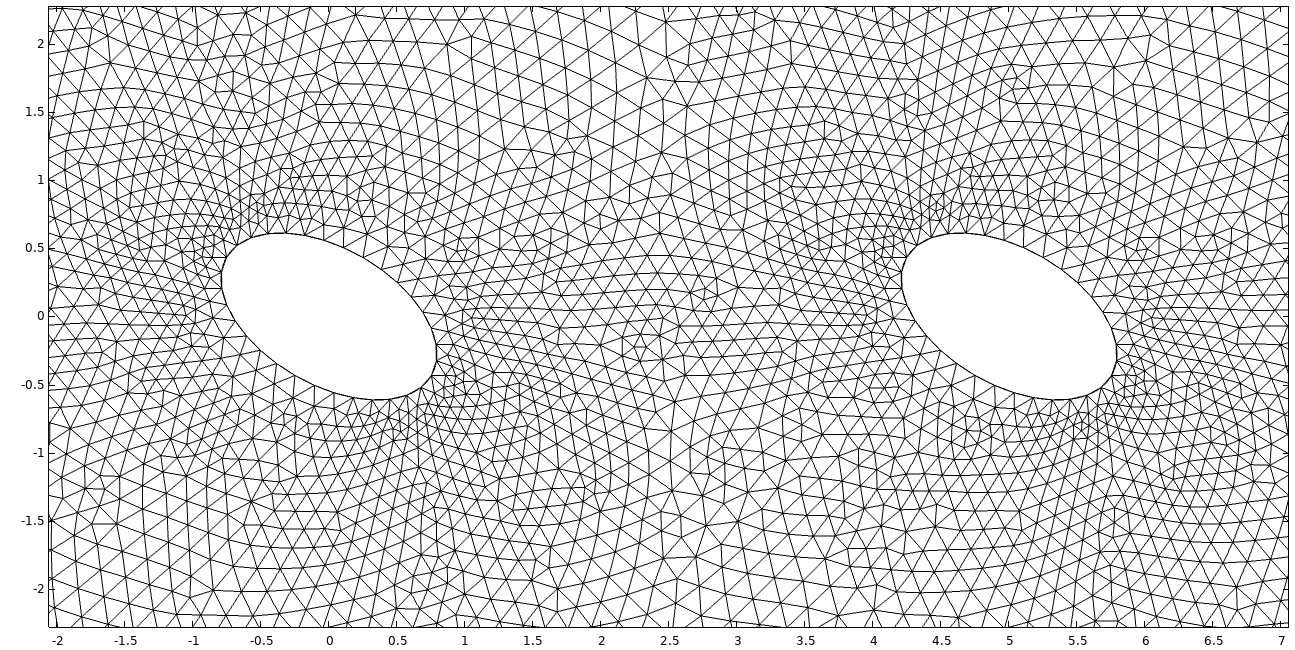}
		\caption{Coarse Mesh (32052 Elements)}
	\end{subfigure}
	\begin{subfigure}[t]{0.32\textwidth}
		\includegraphics[width=\textwidth]{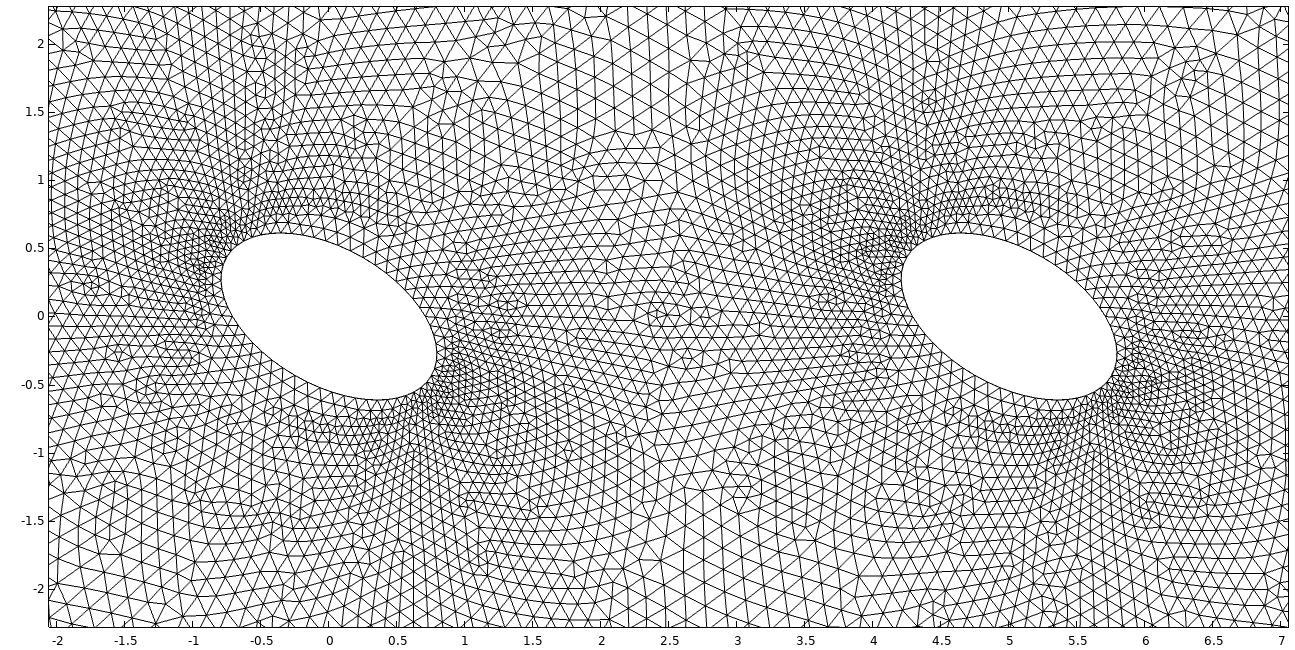}
		\caption{Medium Mesh (118938 Elements)}
	\end{subfigure}
	\begin{subfigure}[t]{0.32\textwidth}
		\includegraphics[width=\textwidth]{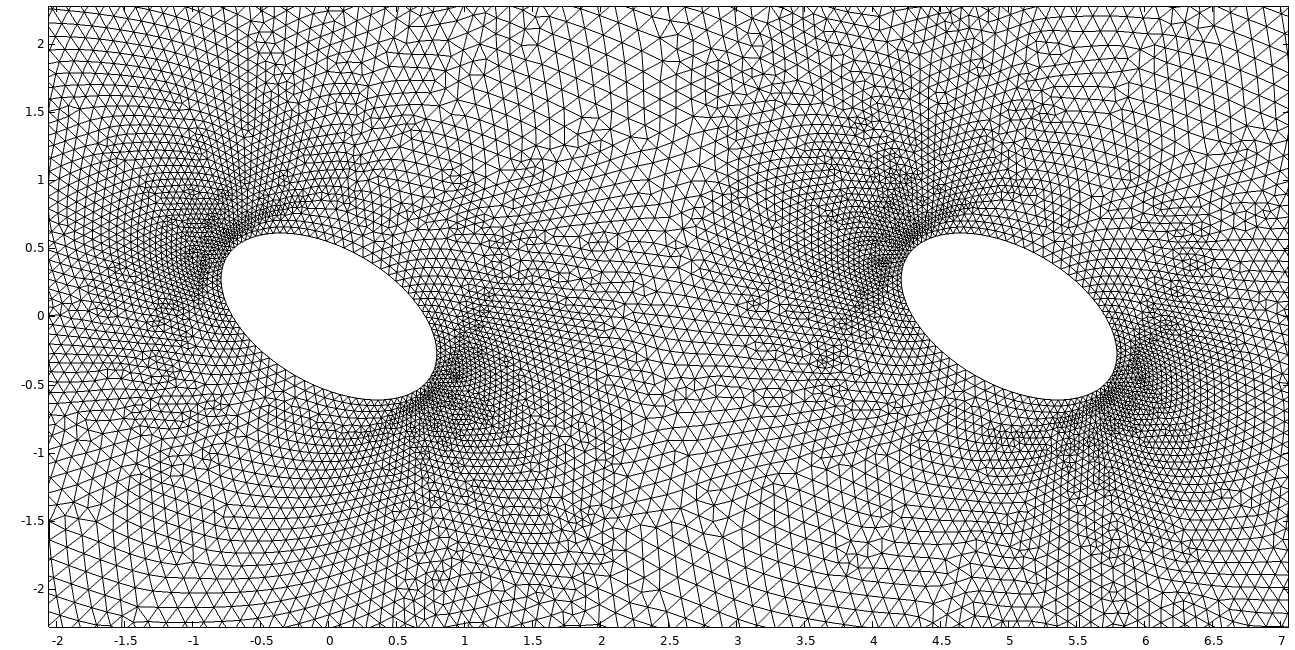}
		\caption{Fine Mesh (457333 Elements)}
	\end{subfigure}
	\caption{Double Cylinder Case: Successively Refined Meshes for Grid-Independence Study}
	\label{Fig: Double Cylinder Case: Grids}
\end{figure}

 The computations are ensured to have low discretization errors by preforming simulations on three different grids, as shown in \cref{Fig: Single Cylinder Case: Grids,Fig: Double Cylinder Case: Grids} for the region around cylinders. The finest grid is used in all the calculations in the following sections. \Cref{Tab: Single Cylinder Case: Lift and drag coefficients for three consecutively refined grids,Tab: Double Cylinder Case: Lift and drag coefficients for three consecutively refined grids} present the computed lift and drag coefficients for the three grids to show the effects of grid refinement for the single and double cylinder cases respectively. The difference between the medium and finest grids is small, and the finest grid can be considered to give grid-independent results.

\begin{table}[H]
	\centering
%	\resizebox{\textwidth}{!}{%
		\begin{tabular}{|c|c|c|}
			\hline
			& Lift & Drag  \\ \hline
			Coarse Mesh & 0.3868        & 1.2160         \\ \hline
			Medium Mesh & 0.3895        & 1.2289        \\ \hline
			Fine Mesh   & 0.3903        & 1.2306        \\ \hline
		\end{tabular}%
%	}
	\caption{Single Cylinder Case: Lift and Drag Coefficients for 3 Refined Grids}
	\label{Tab: Single Cylinder Case: Lift and drag coefficients for three consecutively refined grids}
\end{table}

\begin{table}[H]
	\centering
	\resizebox{\textwidth}{!}{%
		\begin{tabular}{|c|c|c|c|c|}
			\hline
			& Lift Upstream & Drag Upstream & Lift Downstream & Drag Downstream \\ \hline
			Coarse Mesh & 0.3797        & 1.5202        & 0.1239          & 0.5510          \\ \hline
			Medium Mesh & 0.3815        & 1.5341        & 0.1225          & 0.5550          \\ \hline
			Fine Mesh   & 0.3817        & 1.5358        & 0.1215          & 0.5560          \\ \hline
		\end{tabular}%
	}
	\caption{Double Cylinder Case: Lift and Drag Coefficients for 3 Refined Grids}
	\label{Tab: Double Cylinder Case: Lift and drag coefficients for three consecutively refined grids}
\end{table}
%\section{Development of Neural Networks} \label{Sec:Development of Neural Networks}
%\subsection{Description of Neural Networks}
%We have used two different types of neural networks: multilayer perceptron (MLPNN) and convolutional neural network (CNN). Architectures and training processes of both the networks are described in this section.
\section{Multilayer Perceptron Neural Network (MLPNN) for Prediction of Lift and Drag Coefficients} \label{Sec:MLPNN description}
A perceptron, also known as neuron, is a building block of the network. A neuron performs linear transformation on the input vector followed by an element-wise nonlinear activation function to give an output. MLPNN is one of the simplest neural network architectures where neurons are stacked together to form a layer and multiple layers are combined to form a deep network. The linear transformation requires weights and biases which are estimated during the training process by minimizing the loss function. In this work, the mean squared error between estimates obtained from the numerical simulations and predictions of the neural network is defined as the loss function. Hyper-parameters such as number of layers, number of neurons, learning rate etc., are fine tuned by randomly splitting the data into two subsets: training and validation. We use 90\% data for training and 10\% for validation. A smaller network with few hidden neurons and layers does not fit the training data satisfactorily. This phenomenon is known as under-fitting or bias. Adding more neurons and layers increases the nonlinearity and thus, improves the prediction capability of the network. However, excessively deep networks tend to fit the training data with high accuracy but fail to generalize on unseen validation data. This is known as over-fitting or variance. In practice, a network with low bias and low variance is desired. This is achieved by using the validation data. The trained network is tested on an unseen dataset. A well trained network should perform satisfactorily on both training and testing sets i.e., it should demonstrate low errors and high accuracies on both the sets. More details of the network architecture, training procedure, back-propagation algorithm etc. can be found in the literature \cite{goodfellow2016deep, shahane2019numerical}. In this work, we have used the open source Python library TensorFlow \cite{tensorflow2015-whitepaper} with its high level API Keras \cite{chollet2015keras}.
\par
\begin{figure}[H]
	\centering
	\begin{subfigure}[t]{0.44\textwidth}
		\includegraphics[width=\textwidth]{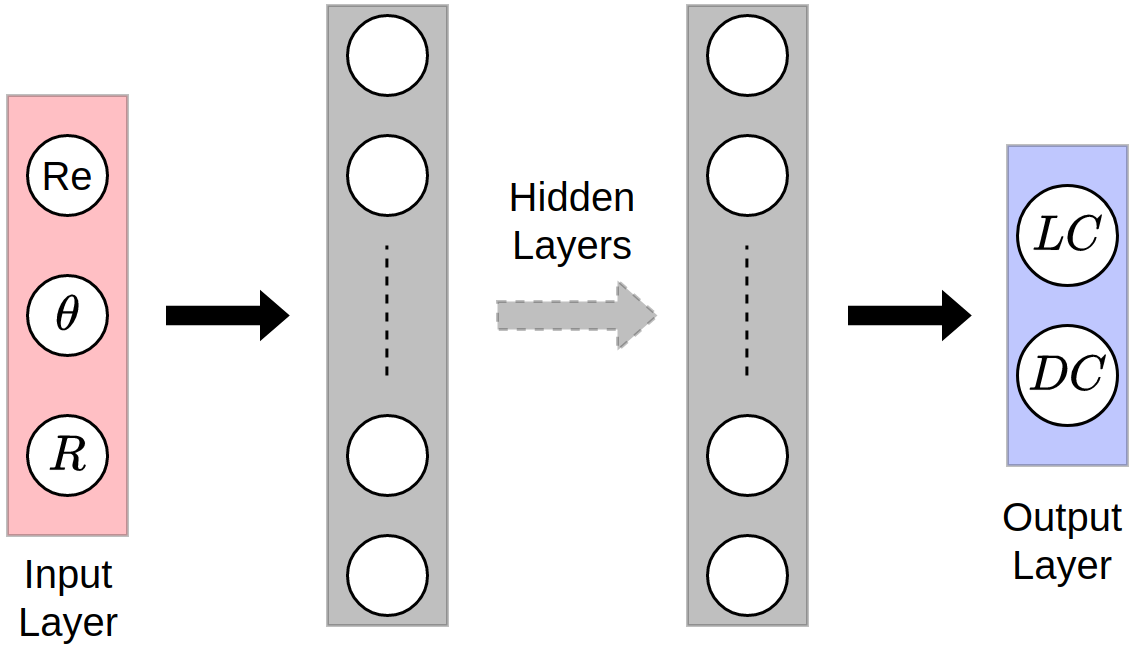}
		\caption{Case of Single Elliptic Cylinder}
		\label{Fig:MLPNN for Estimation of Lift and Drag Coefficients single}
	\end{subfigure}
	\hspace{0.1\textwidth}
	\begin{subfigure}[t]{0.44\textwidth}
		\includegraphics[width=\textwidth]{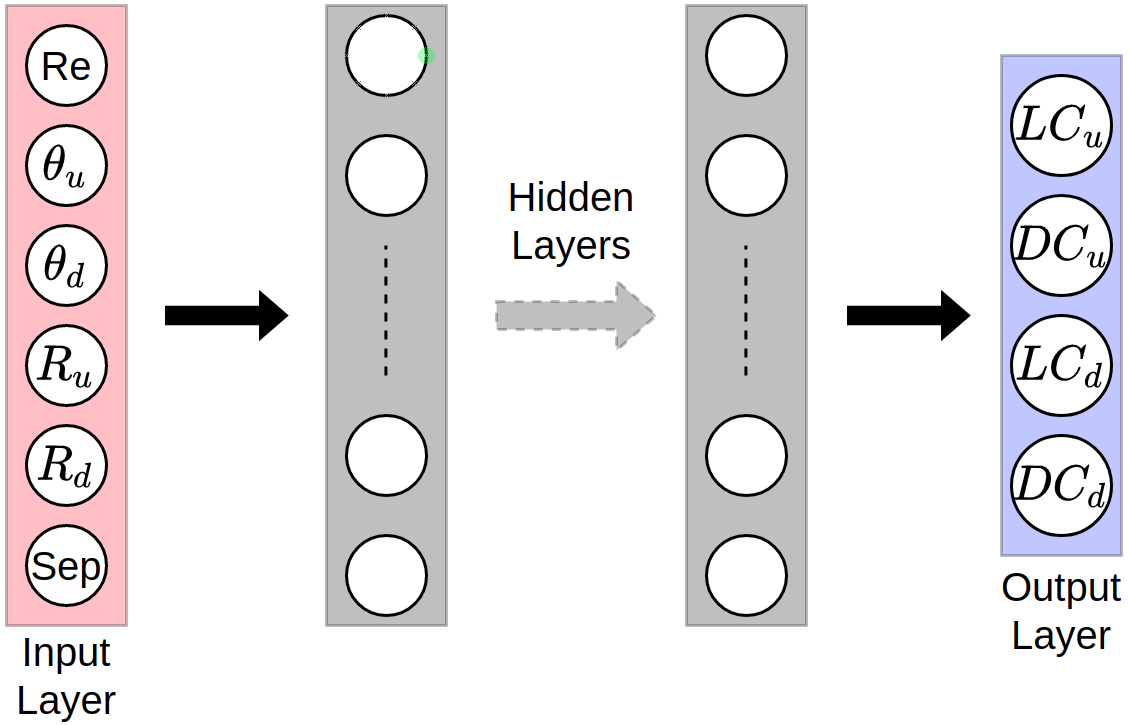}
		\caption{Case of Double Elliptic Cylinder}
		\label{Fig:MLPNN for Estimation of Lift and Drag Coefficients double}
	\end{subfigure}
	\caption{MLPNN for Estimation of Lift and Drag Coefficients ($LC$: lift coefficient, $DC$: drag coefficient, $Re$: Reynolds no., $\theta$: angle of attack, $R$: ratio of major to minor axis, Sep: separation between ellipse centers, subscript $u$: upstream, subscript $d$: downstream)}
	\label{Fig:MLPNN for Estimation of Lift and Drag Coefficients}
\end{figure}

 Lift and drag coefficients are estimated as a function of Reynolds number and geometric parameters using the MLPNN architecture. For the case of a single elliptic cylinder, Reynolds number, angle of attack and ratio of major to minor axis are the three independent input parameters which affect the lift and drag coefficients of the elliptic cylinder (\cref{Fig:MLPNN for Estimation of Lift and Drag Coefficients single}). For the second case of double elliptic cylinder, the separation between them is an additional input parameter together with individual angles of attack and ratios (\cref{Fig:MLPNN for Estimation of Lift and Drag Coefficients double}). The Reynolds number is defined with respect to the upstream elliptic cylinder. The Reynolds number, angle of attack, aspect ratio and separation are varied in the ranges $[20,40]$, $[0^\text{o},180^\text{o}]$, $[1,3]$ and $[4,10]$ respectively. The range of Reynolds number is chosen such that a steady state solution exists for the lowest aspect ratio of unity. For a circular cylinder, the critical Reynolds number is around 42.

\begin{figure}[H]
	\centering
	\begin{subfigure}[t]{0.32\textwidth}
		\includegraphics[width=\textwidth]{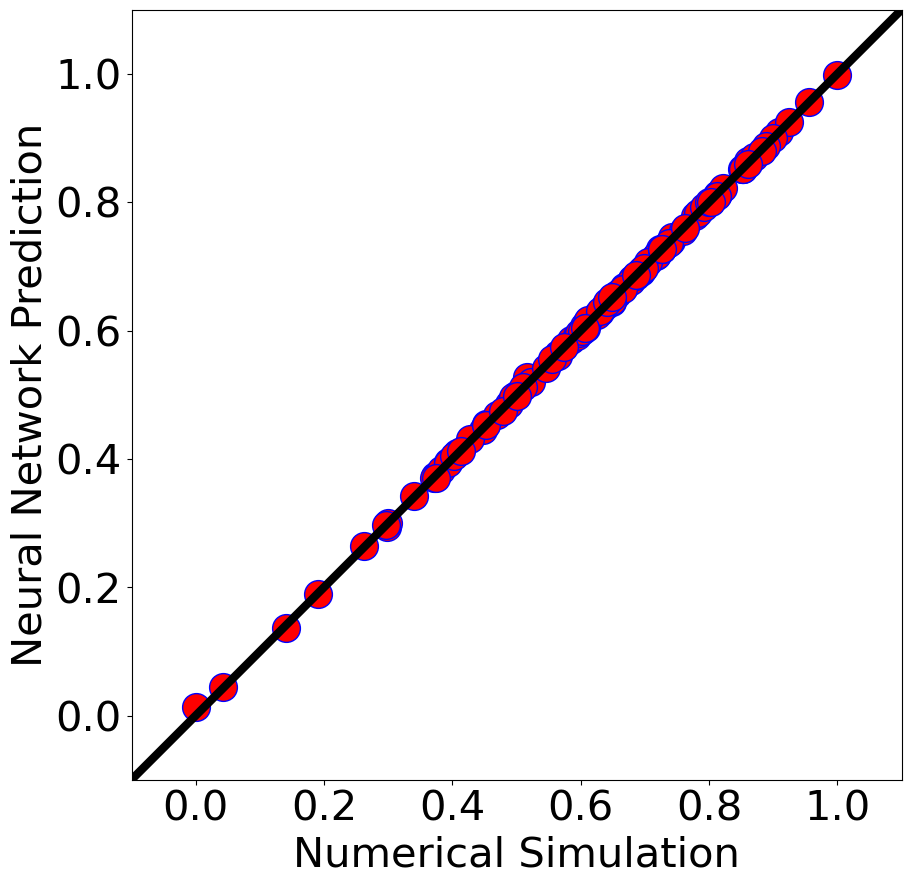}
		\caption{Single Elliptic Cylinder}
	\end{subfigure}
	\begin{subfigure}[t]{0.32\textwidth}
		\includegraphics[width=\textwidth]{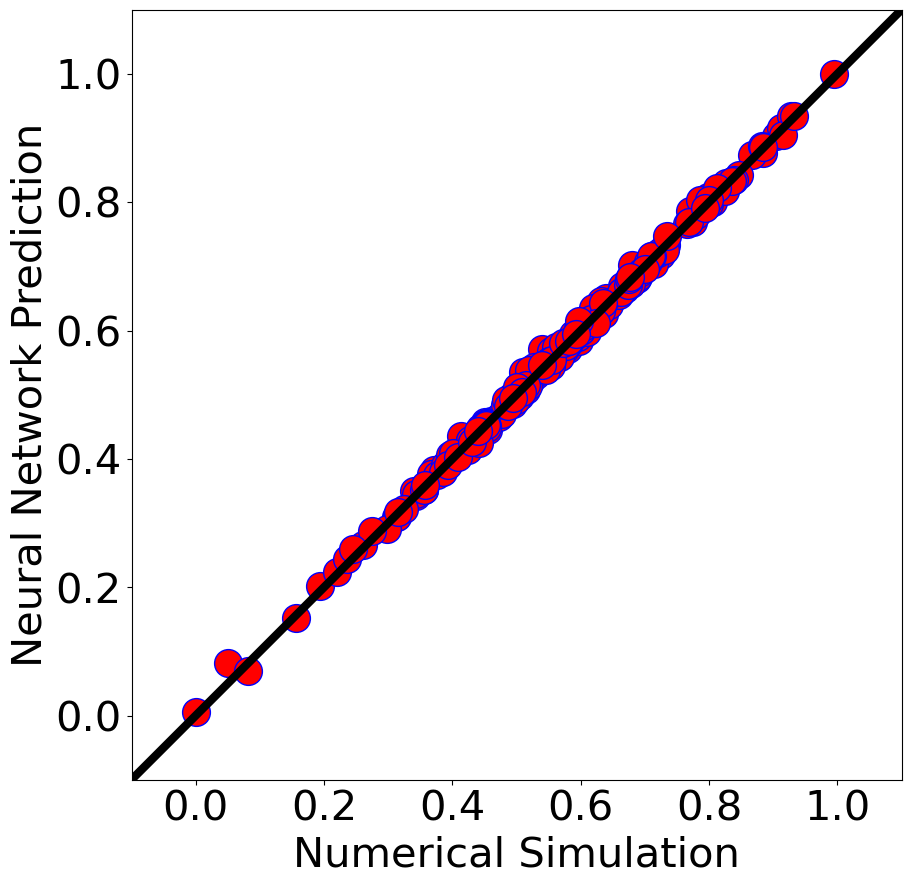}
		\caption{Double Elliptic Cylinder: Upstream}
	\end{subfigure}
	\begin{subfigure}[t]{0.32\textwidth}
		\includegraphics[width=\textwidth]{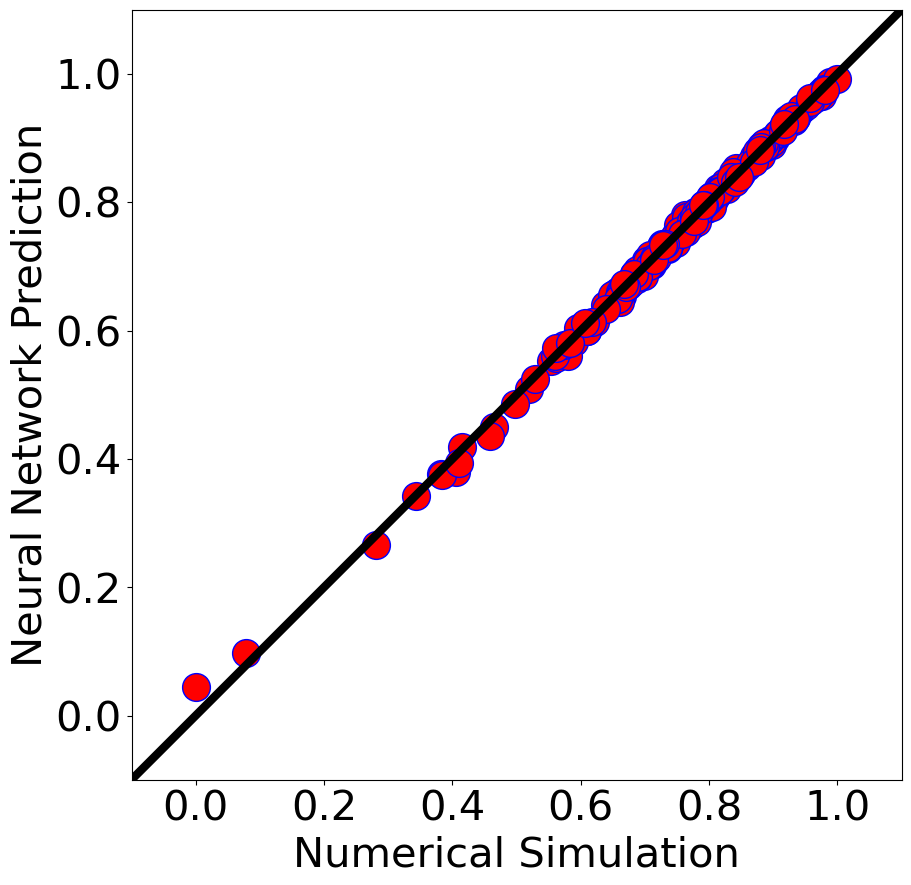}
		\caption{Double Elliptic Cylinder: Downstream}
	\end{subfigure}
	\caption{Drag Coefficient (Scaled): Comparison of Numerical Simulation and Neural Network Prediction}
	\label{Fig:comparison drag}
\end{figure}

\begin{figure}[H]
	\centering
	\begin{subfigure}[t]{0.32\textwidth}
		\includegraphics[width=\textwidth]{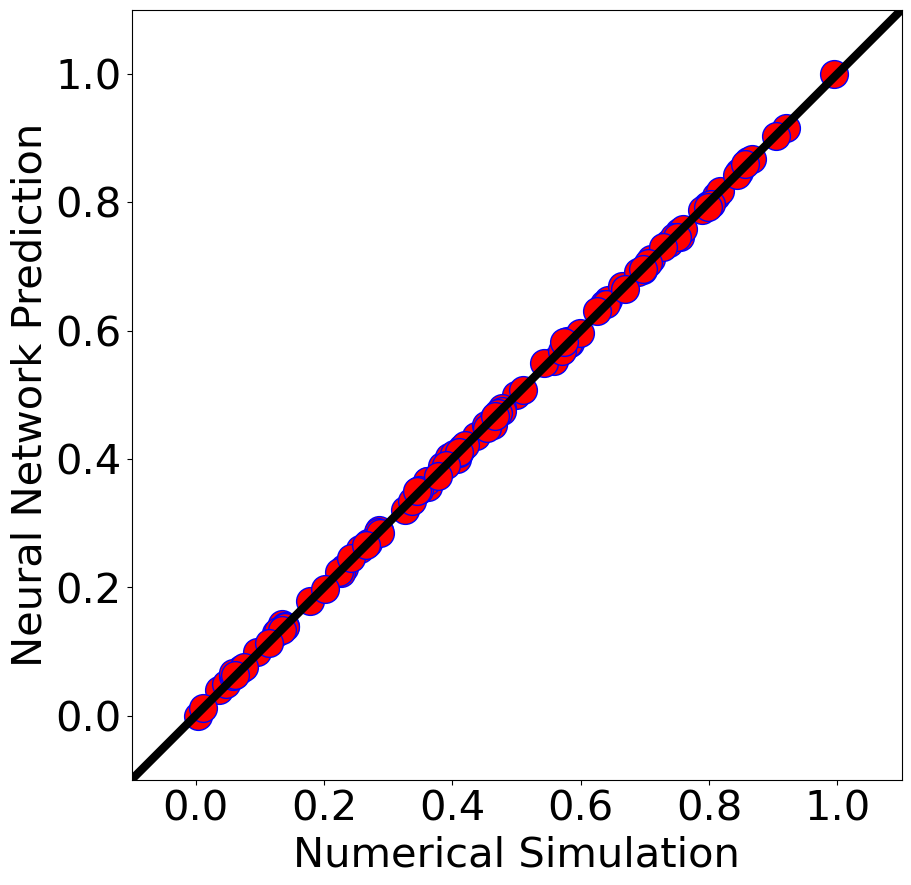}
		\caption{Single Elliptic Cylinder}
	\end{subfigure}
	\begin{subfigure}[t]{0.32\textwidth}
		\includegraphics[width=\textwidth]{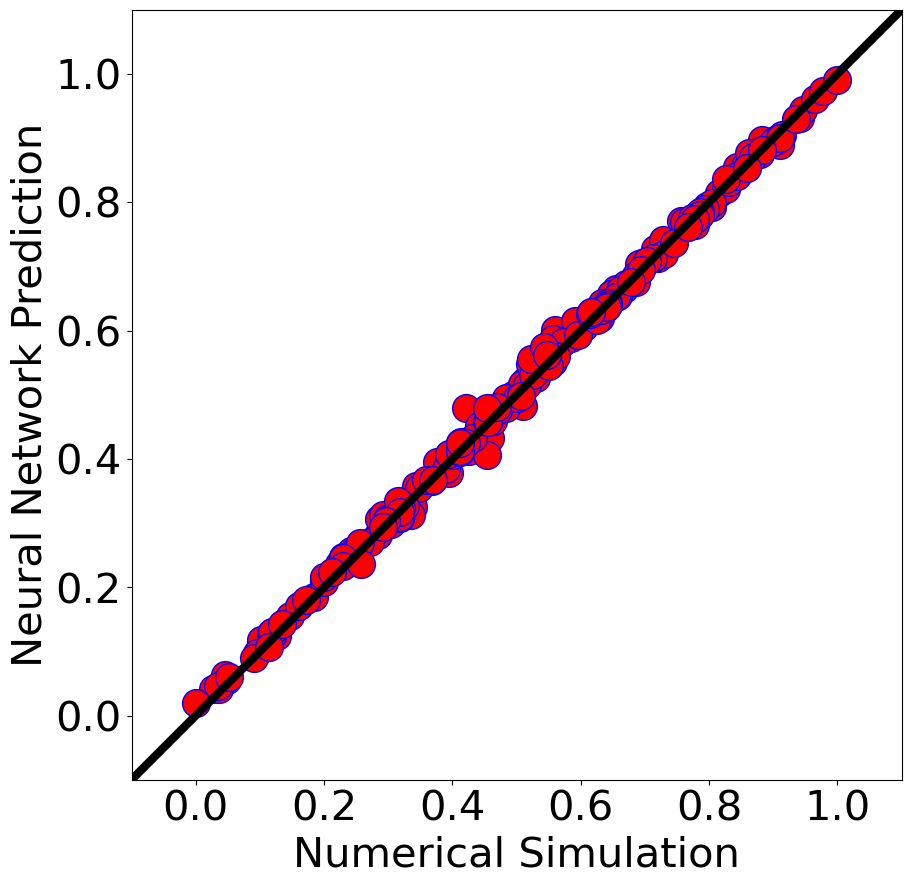}
		\caption{Double Elliptic Cylinder: Upstream}
	\end{subfigure}
	\begin{subfigure}[t]{0.32\textwidth}
		\includegraphics[width=\textwidth]{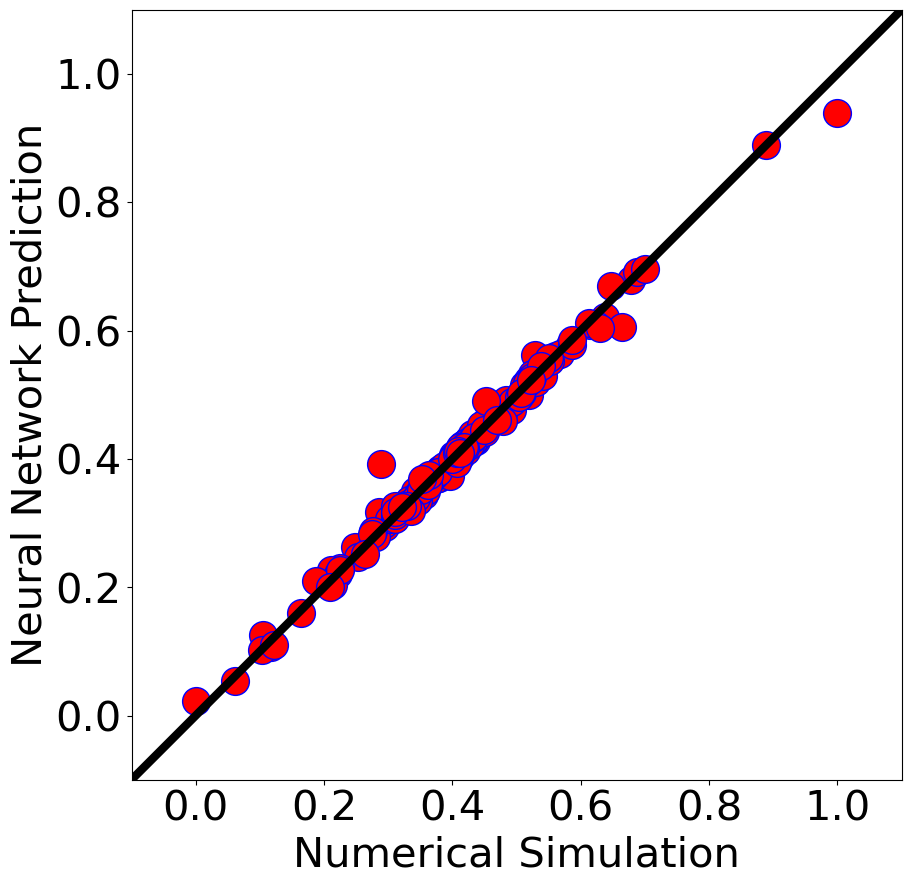}
		\caption{Double Elliptic Cylinder: Downstream}
	\end{subfigure}
	\caption{Lift Coefficient (Scaled): Comparison of Numerical Simulation and Neural Network Prediction}
	\label{Fig:comparison lift}
\end{figure}

\begin{table}[H]
	\centering
	\resizebox{\textwidth}{!}{%
		\begin{tabular}{|c|c|c|}
			\hline
			& Case of Single Elliptic Cylinder     & Case of Double Elliptic Cylinder     \\ \hline
			No. of Hidden Layers            & 6                          & 8                          \\ \hline
			No. of Neurons per Hidden Layer & 100                        & 200                        \\ \hline
			No. of Trainable Parameters     & 51,102                     & 286,804                    \\ \hline
			Learning Rate                   & 0.002                      & 0.1                        \\ \hline
			Optimization Algorithm          & Adam \cite{kingma2014adam} & Adam \cite{kingma2014adam} \\ \hline
			No. of Epochs                   & 1000                       & 10000                      \\ \hline
			Loss Function                   & Mean Squared Error         & Mean Squared Error         \\ \hline
			Hidden Layers Activation        & ReLU                       & ReLU                       \\ \hline
			Output Layer Activation         & Linear                     & Linear                     \\ \hline
			Size of Training Set            & 1100                       & 3500                       \\ \hline
			Size of Testing Set             & 100                        & 200                        \\ \hline
		\end{tabular}%
	}
	\caption{Hyper-Parameters of MLPNN}
	\label{Tab: Hyper-Parameters of MLPNN}
\end{table}
Hyper-parameters of the MLPNN for both the cases are listed in \cref{Tab: Hyper-Parameters of MLPNN}. These hyper-parameters are tuned by using 10\% of the training data for validation. The trained network is tested on a separate unseen dataset. For a dataset with sample size $m$, let $\bm{y}^s=[y_1^s, y_2^s, \dots, y_m^s]$ and $\bm{y}^n=[y_1^n, y_2^n, \dots, y_m^n]$ denote predictions of the variable $y$ using numerical simulations and neural networks respectively. The coefficient of determination ($R^2$) \cite{cameron1997r} is used to estimate accuracy of the neural networks.
\begin{equation}
	\text{Accuracy: } R^2 = 1 - \frac{\sum_{i=1}^{m} (y_i^s - y_i^n)^2}{\sum_{i=1}^{m} (y_i^s - \text{mean}(\bm{y}^s))^2}
	\label{Eq:accuracy}
\end{equation}
Percentage errors are defined as follows:
\begin{equation}
	\text{Average percent error: } 100 \times \frac{1}{m} \frac{\sum_{i=1}^{m} ||y_i^s - y_i^n||}{\max_{i=1}^m ||y_i^s||}
	\label{Eq:average percent error}
\end{equation}
\begin{equation}
	\text{Maximum percent error: } 100 \times \frac{\max_{i=1}^{m} ||y_i^s - y_i^n||}{\max_{i=1}^m ||y_i^s||}
	\label{Eq:maximum percent error}
\end{equation}
\begin{table}[H]
	\centering
	\resizebox{\textwidth}{!}{%
		\begin{tabular}{|c|c|c c|c c|c c|}
			\hline
			\multicolumn{2}{|c|}{\multirow{3}{*}{}} & \multicolumn{2}{c|}{\multirow{2}{*}{\begin{tabular}[c]{@{}c@{}}Single \\ Elliptic Cylinder\end{tabular}}} & \multicolumn{4}{c|}{Double Elliptic Cylinder} \\ \cline{5-8}
			\multicolumn{2}{|c|}{} & \multicolumn{2}{c|}{} & \multicolumn{2}{c|}{Upstream} & \multicolumn{2}{c|}{Downstream} \\ \cline{3-8}
			\multicolumn{2}{|c|}{} & Lift & Drag & Lift & Drag & Lift & Drag \\ \hline
			\multirow{2}{*}{Accuracy} & Training & 0.999936 & 0.999954 & 0.998668 & 0.998507 & 0.997685 & 0.998775 \\ %\cline{2-8}
			& Testing & 0.999791 & 0.999821 & 0.997543 & 0.998020 & 0.988764 & 0.997606 \\ \hline
			\multirow{2}{*}{\begin{tabular}[c]{@{}c@{}}Average \\ Percent Error\end{tabular}} & Training & 0.3050 & 0.0717 & 1.317 & 0.4091 & 0.6759 & 0.4286 \\ %\cline{2-8}
			& Testing & 0.5495 & 0.1333 & 1.699 & 0.4508 & 1.239 & 0.6132 \\ \hline
			\multirow{2}{*}{\begin{tabular}[c]{@{}c@{}}Maximum \\ Percent Error\end{tabular}} & Training & 1.327 & 0.3679 & 6.807 & 6.235 & 6.706 & 2.609 \\ %\cline{2-8}
			& Testing & 2.063 & 0.9228 & 11.18 & 2.406 & 17.90 & 4.748 \\ \hline
		\end{tabular}%
	}
	\caption{Accuracy and Error of MLPNN}
	\label{Tab: Accuracy and Error of MLPNN}
\end{table}
The error and accuracy for training and testing sets are listed in \cref{Tab: Accuracy and Error of MLPNN}. For a perfect model which fits the data exactly, $R^2$ takes a value of unity \cite{cameron1997r}. In the case of practical models, $R^2$ is always less than unity. $R^2$ value close to unity indicates high accuracy. Thus, low errors and high accuracy show that the networks are successful in estimating the lift and drag coefficients for both the cases. Moreover, similar error estimates for training and testing sets indicate that the chosen hyper-parameters are optimal and the networks have low variance.  The estimates of drag and lift coefficients obtained from numerical simulations and neural networks for testing datasets are plotted in \cref{Fig:comparison drag,Fig:comparison lift}. Both the axes are scaled to range $[0,1]$ using the minimum and maximum values of the numerical estimates. It can be seen that most of the points lie on the ideal trend line $y=x$ indicating high accuracy of the networks. It should be noted that there are always a few outliers which show slightly higher values of maximum percent errors in \cref{Tab: Accuracy and Error of MLPNN}.

\section{Variation of Lift and Drag Coefficients}\label{Sec:Variation of Lift and Drag Coefficients}
\begin{figure}[H]
	\centering
%	\begin{subfigure}[t]{0.49\textwidth}
%		\includegraphics[width=\textwidth]{Figures/single_ellipse/lift_drag/lift_vs_angle_ratio_Re_10.png}
%		\caption{Reynolds No.: 10}
%	\end{subfigure}
	\begin{subfigure}[t]{0.32\textwidth}
		\includegraphics[width=\textwidth]{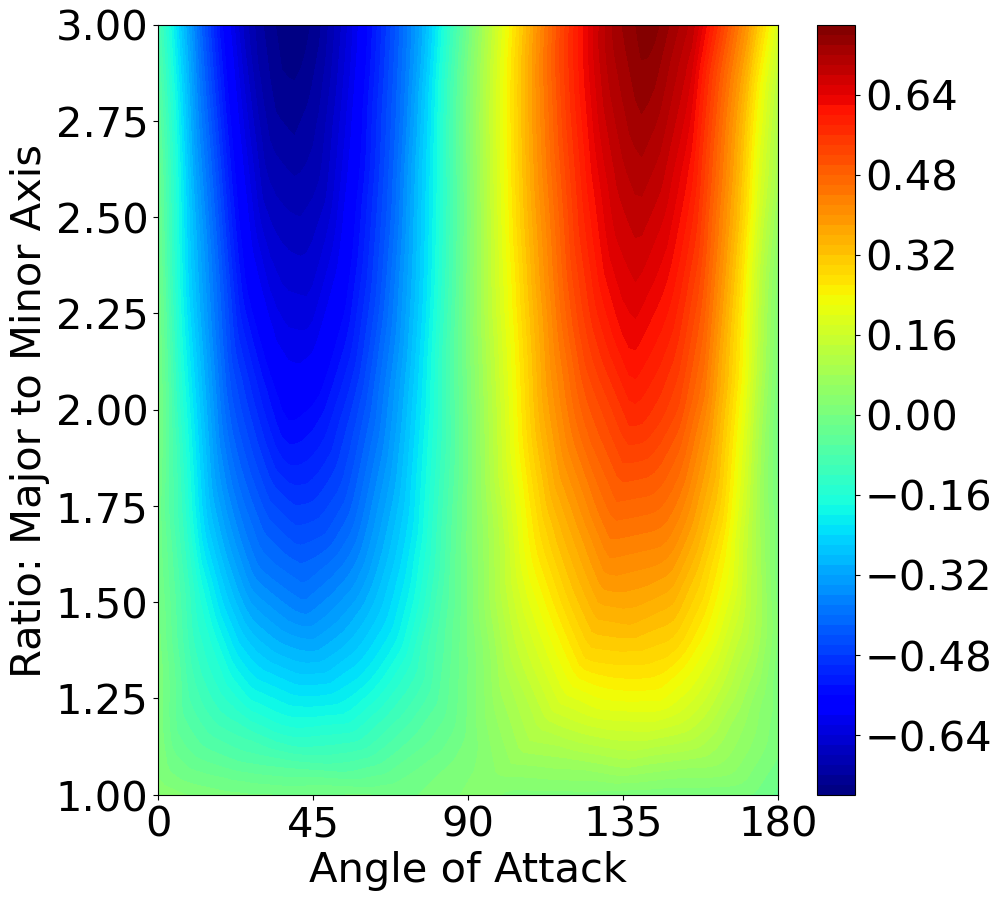}
		\caption{Reynolds No.: 20}
	\end{subfigure}
	\begin{subfigure}[t]{0.32\textwidth}
		\includegraphics[width=\textwidth]{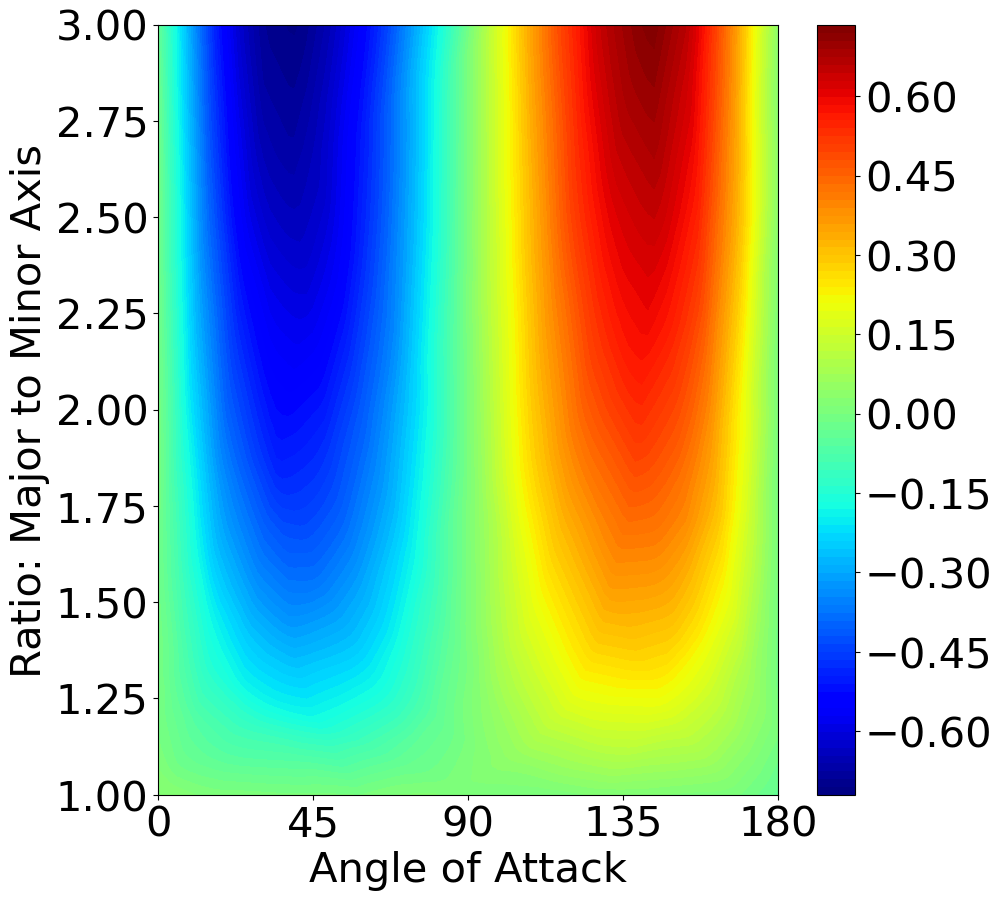}
		\caption{Reynolds No.: 30}
	\end{subfigure}
	\begin{subfigure}[t]{0.32\textwidth}
		\includegraphics[width=\textwidth]{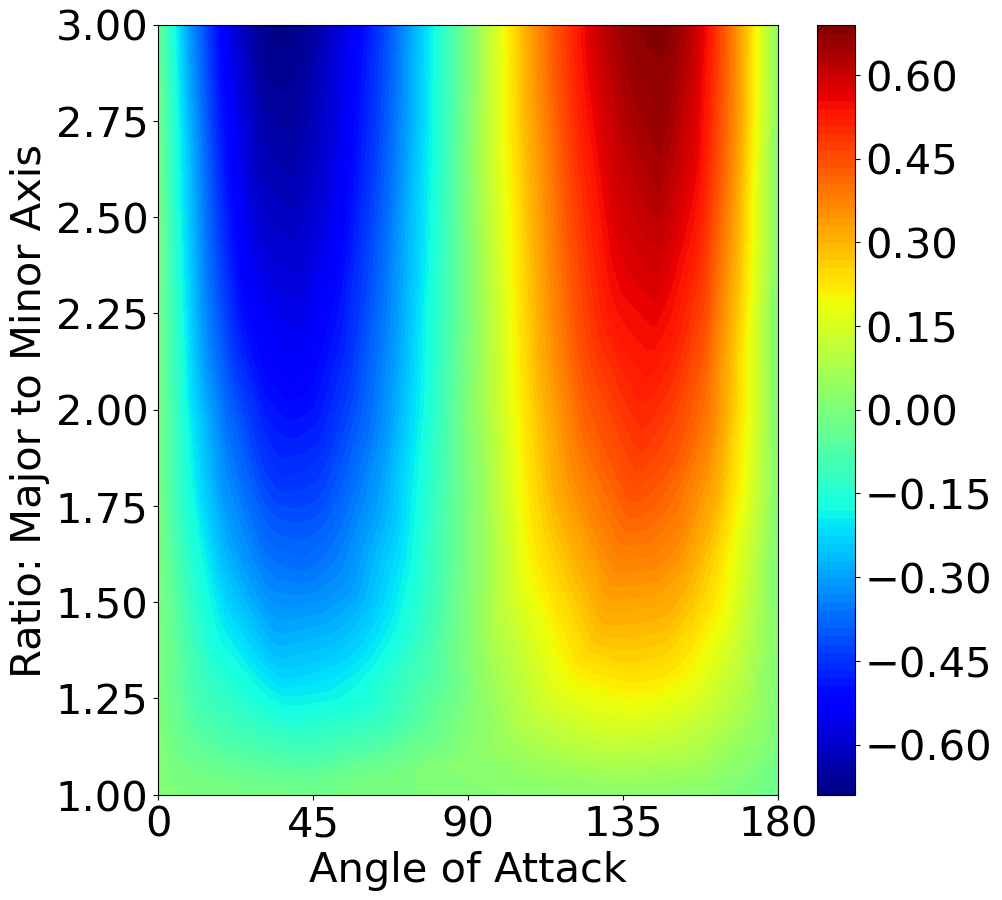}
		\caption{Reynolds No.: 40}
	\end{subfigure}
	\caption{Case of Single Elliptic Cylinder: Lift Coefficient}
	\label{Fig:single ellipse lift vs ratio_angle}
\end{figure}

\begin{figure}[H]
	\centering
%	\begin{subfigure}[t]{0.49\textwidth}
%		\includegraphics[width=\textwidth]{Figures/single_ellipse/lift_drag/drag_vs_angle_ratio_Re_10.png}
%		\caption{Reynolds No.: 10}
%	\end{subfigure}
	\begin{subfigure}[t]{0.32\textwidth}
		\includegraphics[width=\textwidth]{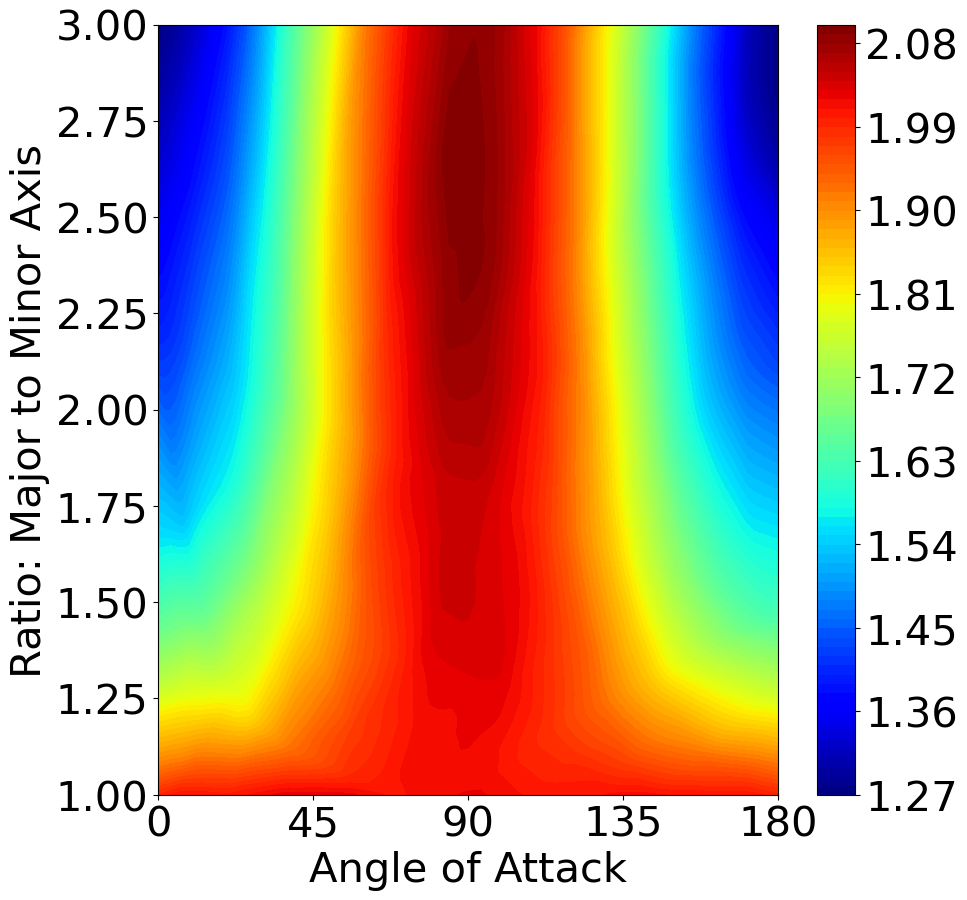}
		\caption{Reynolds No.: 20}
		\label{Fig:single ellipse drag vs ratio_angle Re 20}
	\end{subfigure}
	\begin{subfigure}[t]{0.32\textwidth}
		\includegraphics[width=\textwidth]{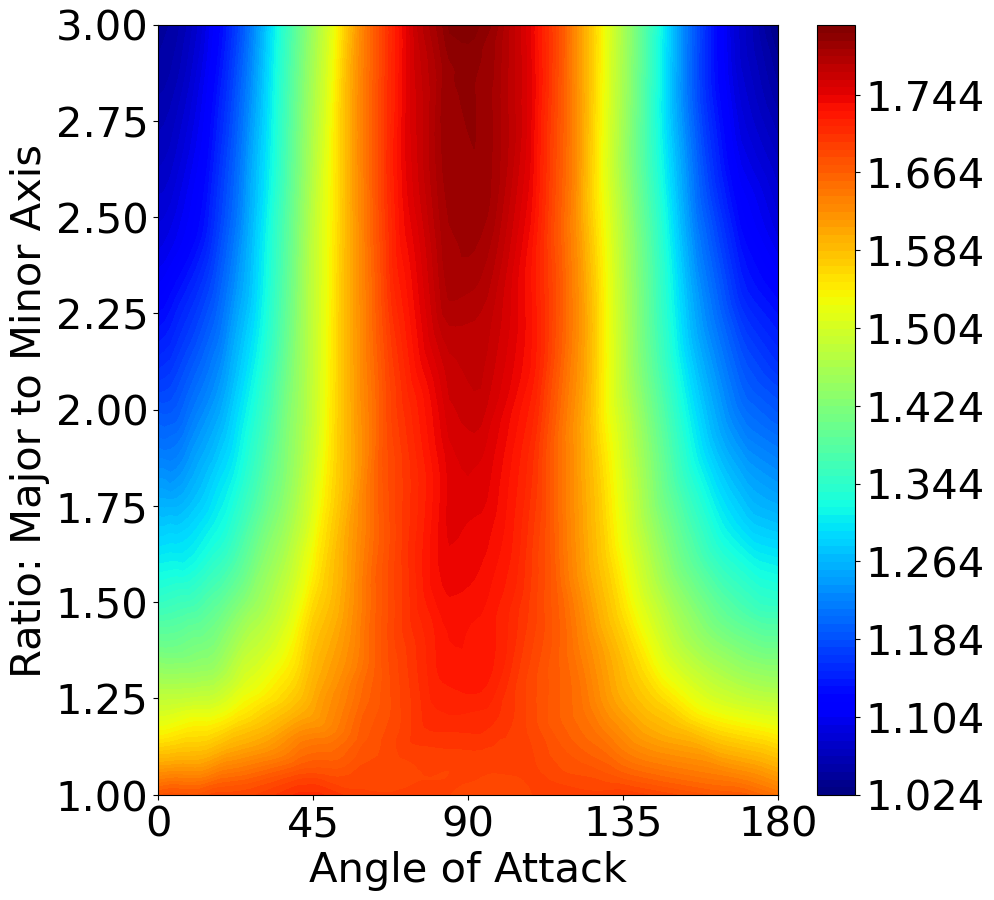}
		\caption{Reynolds No.: 30}
	\end{subfigure}
	\begin{subfigure}[t]{0.32\textwidth}
		\includegraphics[width=\textwidth]{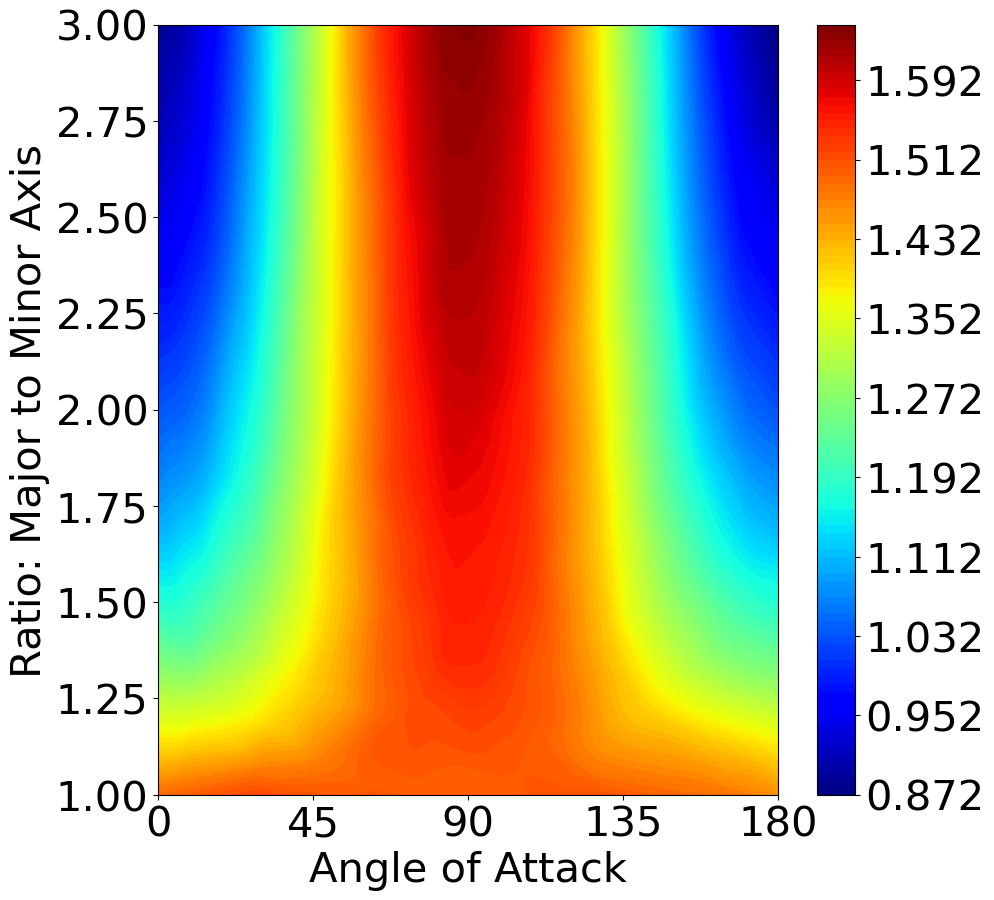}
		\caption{Reynolds No.: 40}
	\end{subfigure}
	\caption{Case of Single Elliptic Cylinder: Drag Coefficient}
	\label{Fig:single ellipse drag vs ratio_angle}
\end{figure}

The MLPNN is now used to study the variation of lift and drag coefficients with aspect ratio, angle of attack and the flow Reynolds number. A pictorial variation as contour plots is first shown in \cref{Fig:single ellipse lift vs ratio_angle,Fig:single ellipse drag vs ratio_angle}. \Cref{Fig:single ellipse lift vs ratio_angle} shows the lift coefficients with angle of attack and aspect ratio as coordinate axes. The angle of $0^{\text{o}}$ corresponds to a horizontal placement of the major axis, and the angle of attack is measured counterclockwise. An angle of $90^{\text{o}}$ corresponds to the case of major axis being vertical. As can be expected, the lift coefficients for $0^{\text{o}}$, $90^{\text{o}}$, and $180^{\text{o}}$ angles of attack are zero, with maximum lift coefficients occurring around $45^{\text{o}}$ and $135^{\text{o}}$. Moreover, for angle of attack of between $[0^{\text{o}},90^{\text{o}}]$, the major axis points in the first and third quadrant. Hence, the lift is negative since the force is acting in the downward direction. %The variations for Reynolds numbers of 20 and 40 are similar, but their magnitudes are different.
\Cref{Fig:single ellipse drag vs ratio_angle} shows the variation of the drag coefficient at the same three Reynolds numbers. The drag coefficient is always maximum at 90 degrees. The drag is the same for all angles of attack for the aspect ratio of unity (circular cylinder). For non-unity aspect ratios, the drag increases with angle of attack between $0^{\text{o}}$ and $90^{\text{o}}$, and then decreases between $90^{\text{o}}$ and $180^{\text{o}}$. The drag also increases with aspect ratio for any angle of attack. The drag, which includes both pressure drag and viscous shear stress, decreases with Reynolds number because of the lower viscosity. The present Reynolds number range is limited to a value for which all angles of attack give a steady flow.

\begin{figure}[H]
	\centering
	\begin{subfigure}[t]{0.32\textwidth}
		\includegraphics[width=\textwidth]{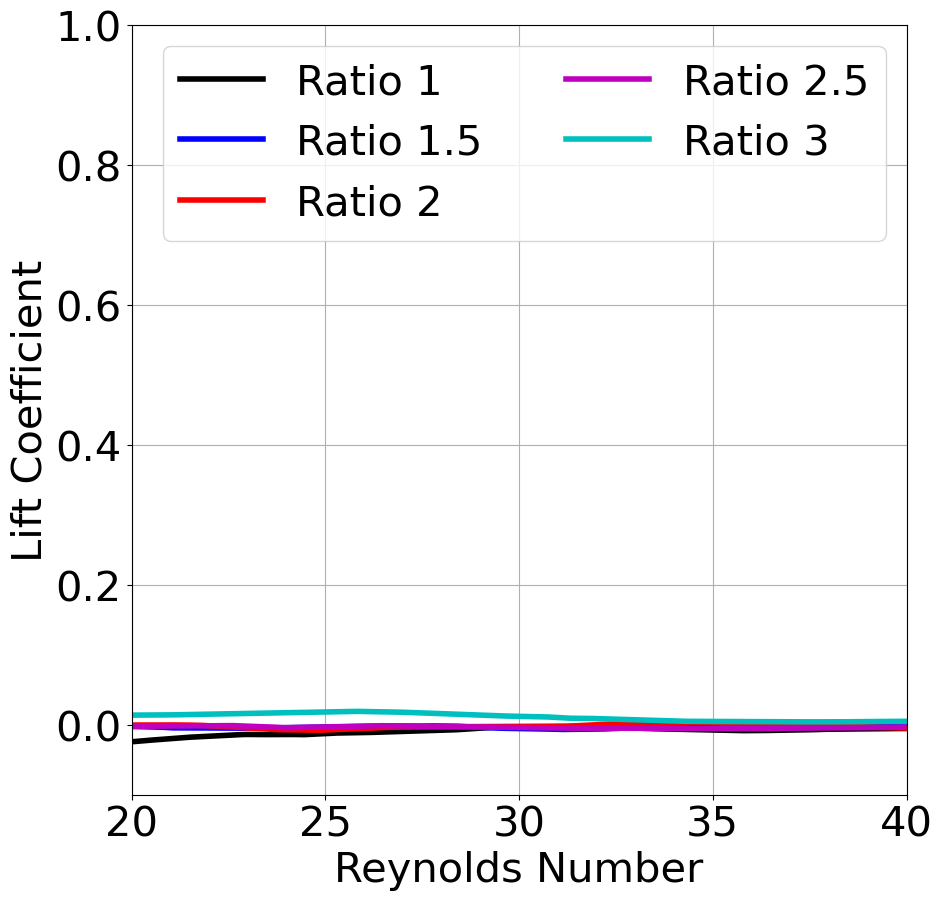}
		\caption{Angle of Attack: $90^0$}
	\end{subfigure}
	\begin{subfigure}[t]{0.32\textwidth}
		\includegraphics[width=\textwidth]{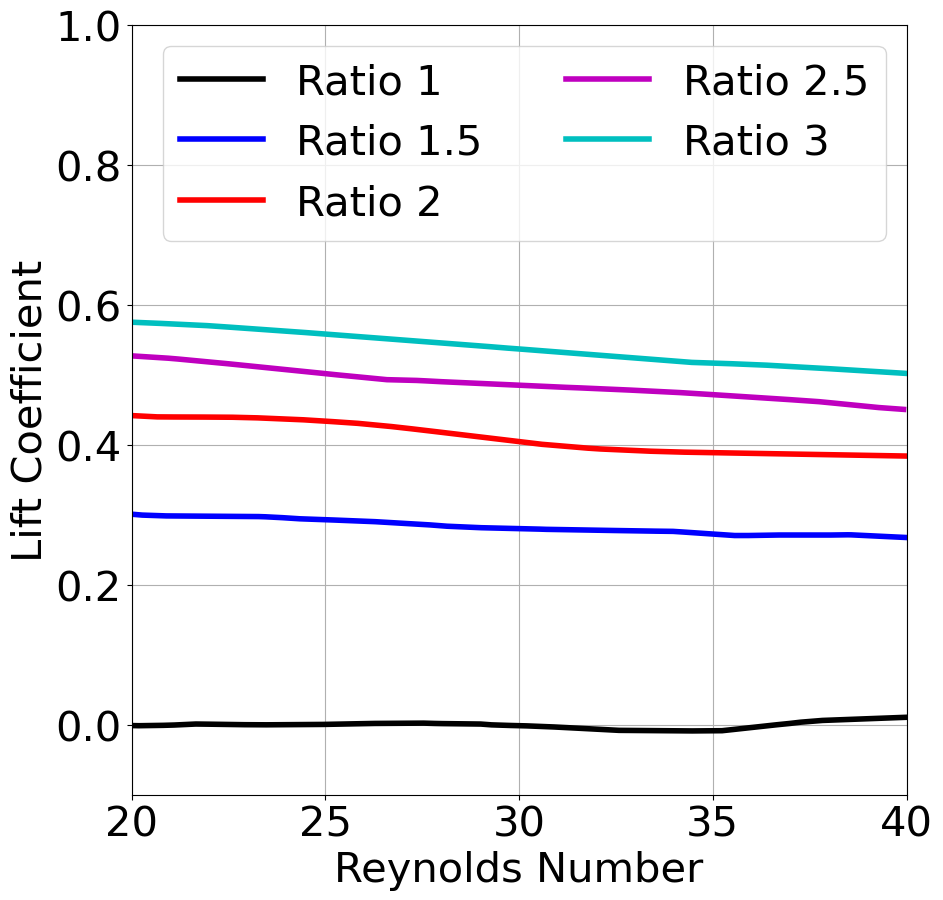}
		\caption{Angle of Attack: $120^0$}
	\end{subfigure}
	\begin{subfigure}[t]{0.32\textwidth}
		\includegraphics[width=\textwidth]{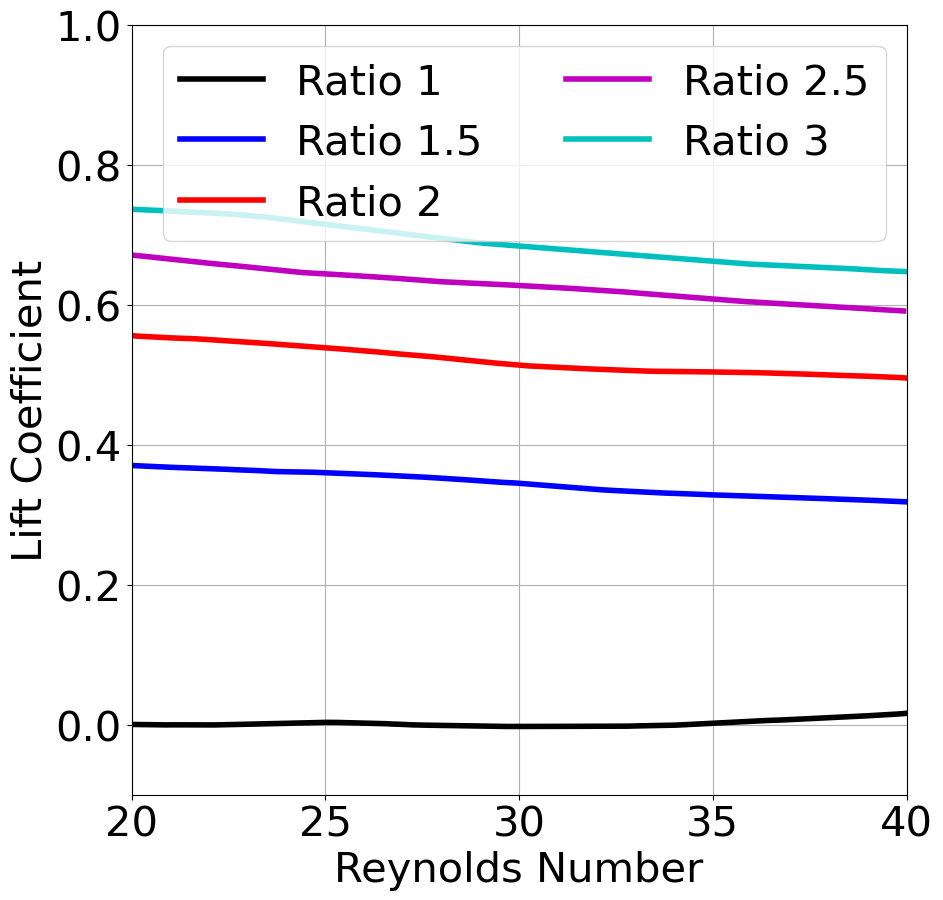}
		\caption{Angle of Attack: $135^0$}
	\end{subfigure}
	\begin{subfigure}[t]{0.32\textwidth}
		\includegraphics[width=\textwidth]{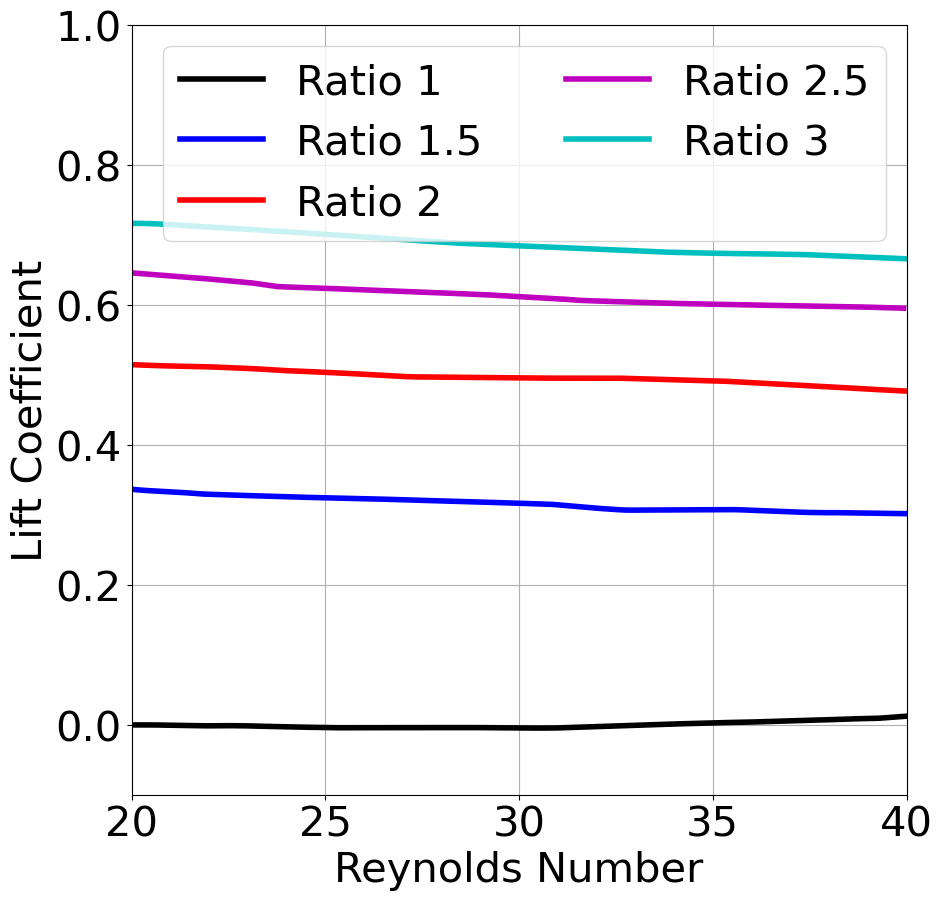}
		\caption{Angle of Attack: $150^0$}
	\end{subfigure}
	\begin{subfigure}[t]{0.32\textwidth}
		\includegraphics[width=\textwidth]{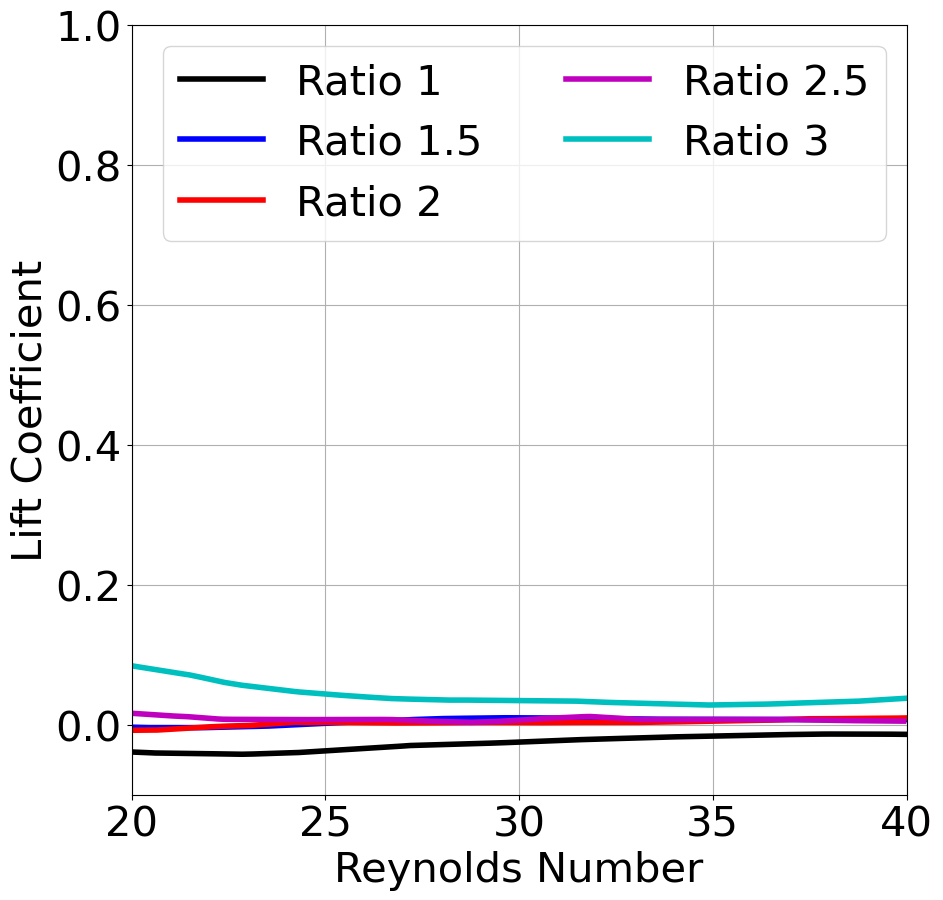}
		\caption{Angle of Attack: $180^0$}
	\end{subfigure}
	\caption{Case of Single Elliptic Cylinder: Lift Coefficient}
	\label{Fig:single ellipse lift vs reynolds varying angle}
\end{figure}
We now discuss the variations in lift and drag coefficients by studying the trends through line plots. \Cref{Fig:single ellipse lift vs reynolds varying angle} shows the lift coefficient for various angles of attack as a function of the Reynolds number. We first observe that the neural network gives the lift coefficient for angles of attack of $0^{\text{o}}$, $90^{\text{o}}$ and $180^{\text{o}}$ to be nearly zero within a few percent error. For other angles between $90^{\text{o}}$ and $180^{\text{o}}$, the lift coefficient is positive, and increases with aspect ratio. As the body becomes slimmer and slimmer, the separation length and the low pressure on the back side of the body increase, thus increasing the lift coefficient. Also, the surface area over which the pressure and normal stress forces act also increases. Thus, the lift coefficient becomes larger as the angle of attack passes $90^{\text{o}}$. The effect of the aspect ratio is always seen to be monotonic for all angles of attack.
\begin{figure}[H]
	\centering
	\begin{subfigure}[t]{0.32\textwidth}
		\includegraphics[width=\textwidth]{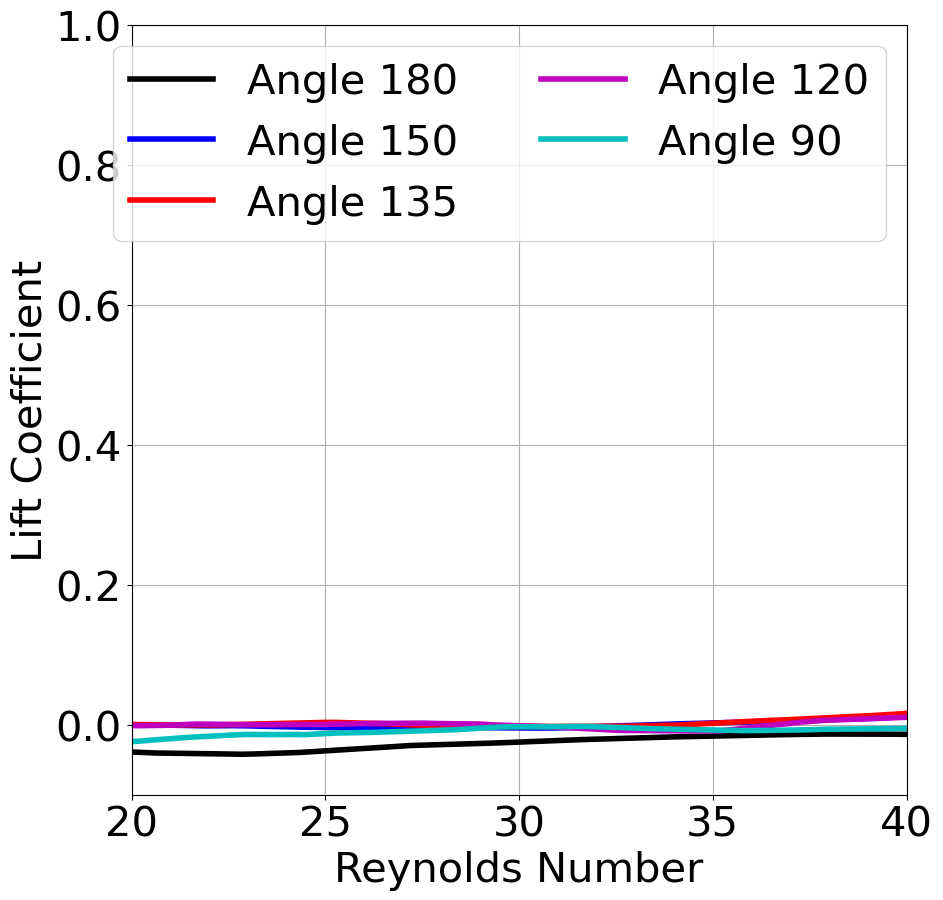}
		\caption{Ratio: Major/Minor Axis: 1}
	\end{subfigure}
	\begin{subfigure}[t]{0.32\textwidth}
		\includegraphics[width=\textwidth]{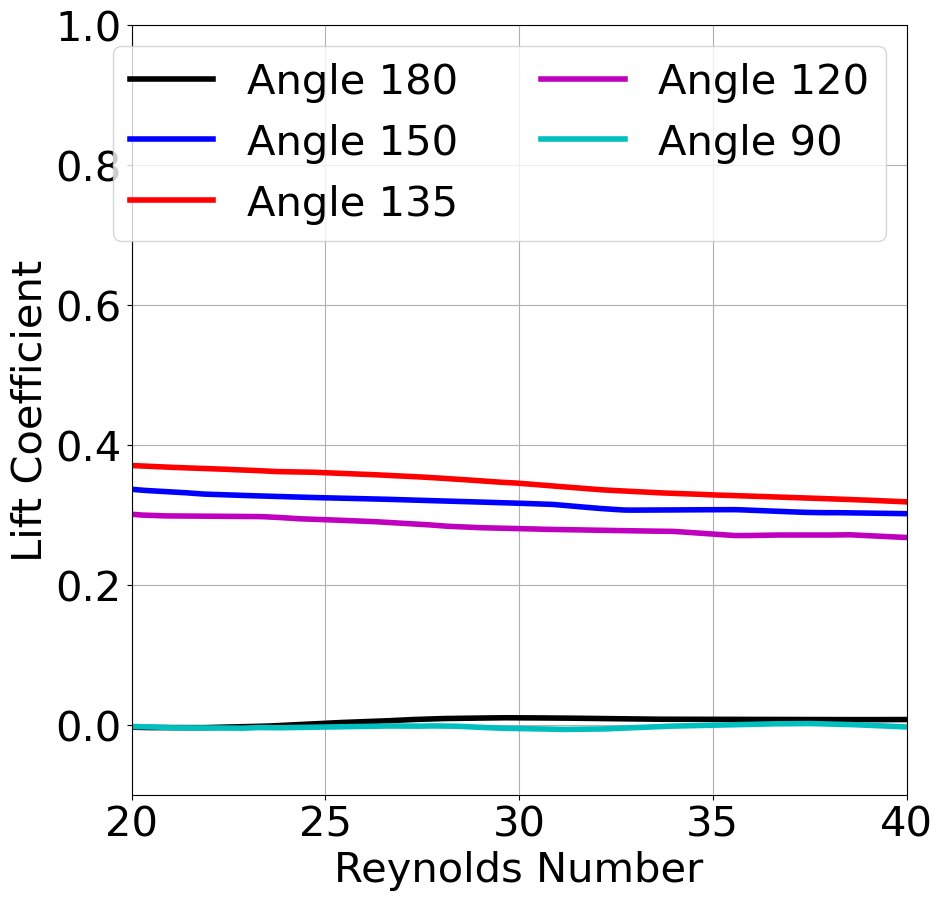}
		\caption{Ratio: Major/Minor Axis: 1.5}
	\end{subfigure}
	\begin{subfigure}[t]{0.32\textwidth}
		\includegraphics[width=\textwidth]{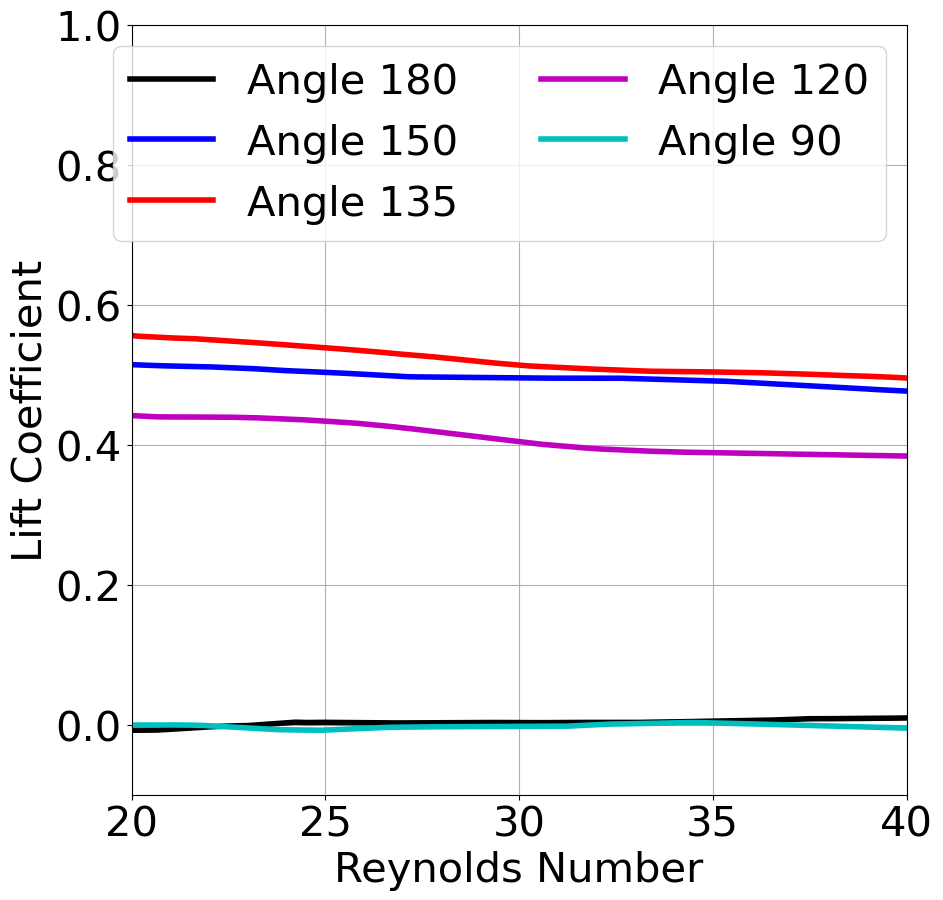}
		\caption{Ratio: Major/Minor Axis: 2}
	\end{subfigure}
	\begin{subfigure}[t]{0.32\textwidth}
		\includegraphics[width=\textwidth]{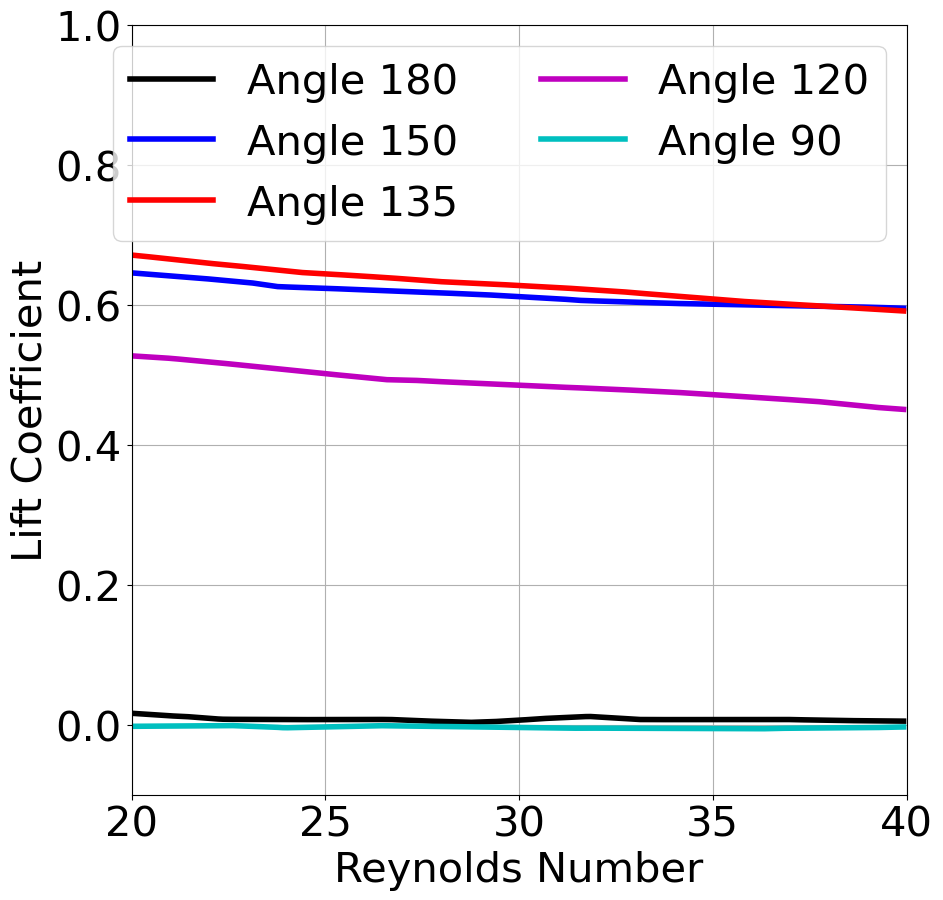}
		\caption{Ratio: Major/Minor Axis: 2.5}
	\end{subfigure}
	\begin{subfigure}[t]{0.32\textwidth}
		\includegraphics[width=\textwidth]{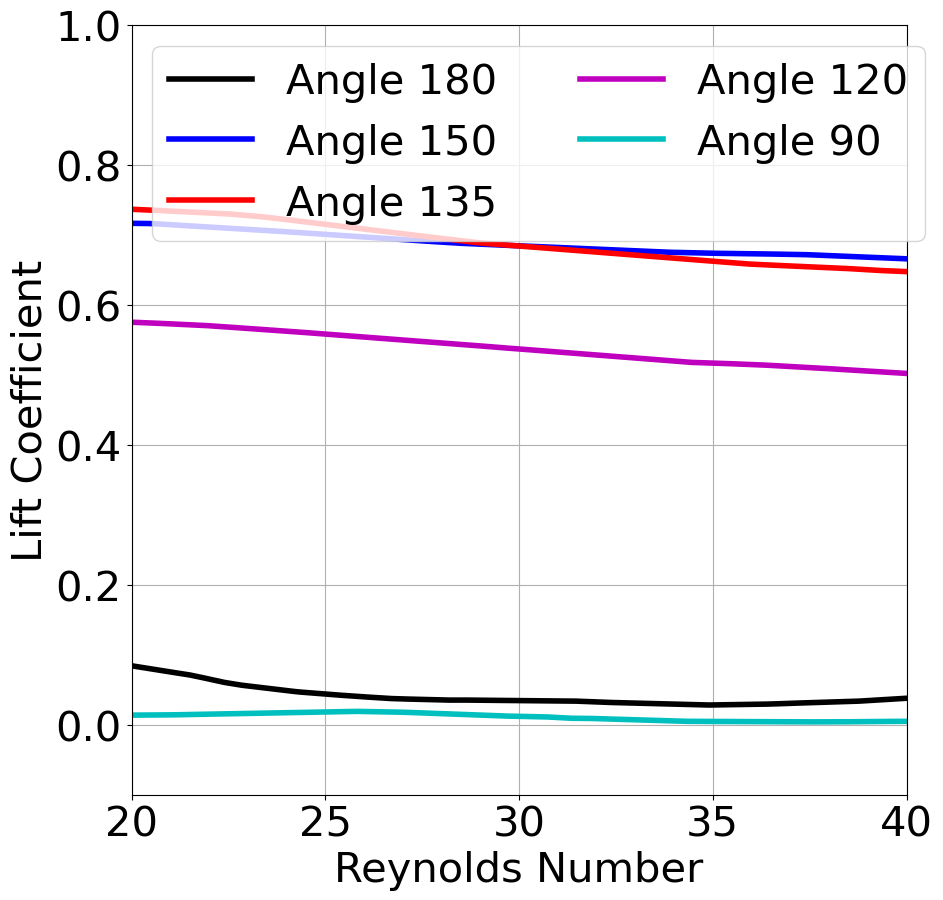}
		\caption{Ratio of Major/Minor Axis: 3}
	\end{subfigure}
	\caption{Case of Single Elliptic Cylinder: Lift Coefficient}
	\label{Fig:single ellipse lift vs reynolds varying ratio}
\end{figure}
It is also instructive to plot the variations of the lift coefficient as a function of the Reynolds number. This effect is seen to be somewhat weak, as negative pressure on the back side of the cylinder decreases only slightly with the Reynolds number. However, as seen earlier, the angle of attack has a predominant effect on the lift coefficient. Surprisingly, the lift coefficients for angle of attack of $150^{\text{o}}$ and $135^{\text{o}}$ are nearly the same for all aspect ratios. This is particularly the case for aspect ratios of two and greater.
\begin{figure}[H]
	\centering
	\begin{subfigure}[t]{0.32\textwidth}
		\includegraphics[width=\textwidth]{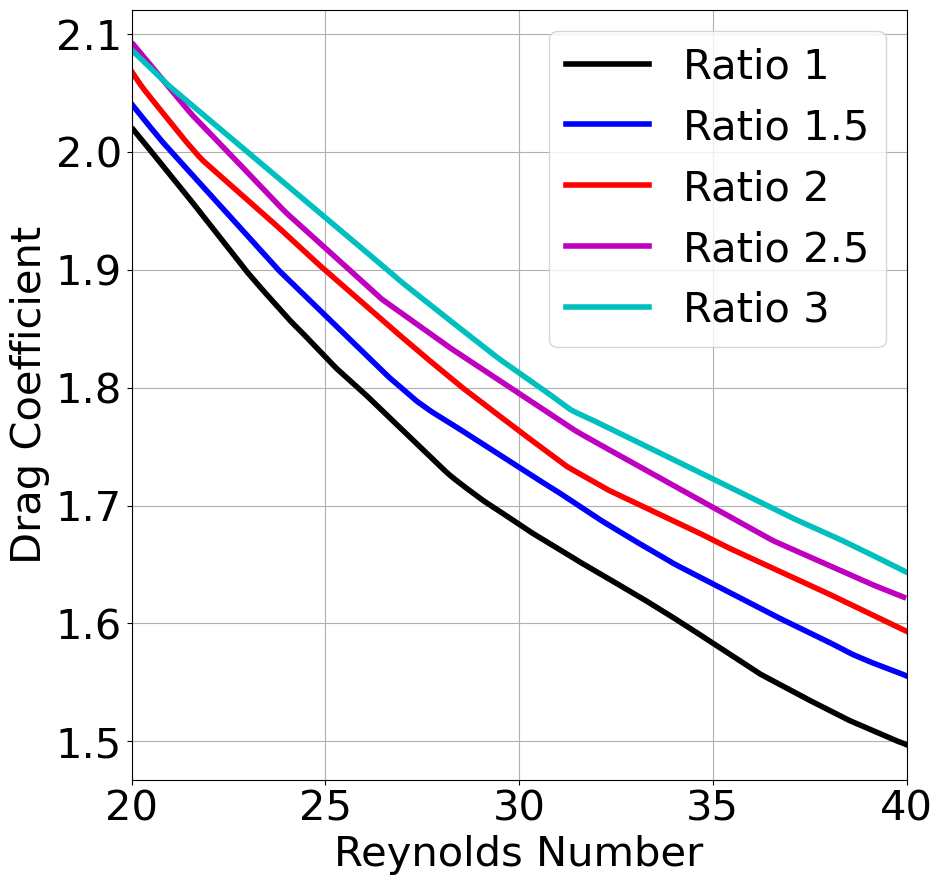}
		\caption{Angle of Attack: $90^0$}
	\end{subfigure}
	\begin{subfigure}[t]{0.32\textwidth}
		\includegraphics[width=\textwidth]{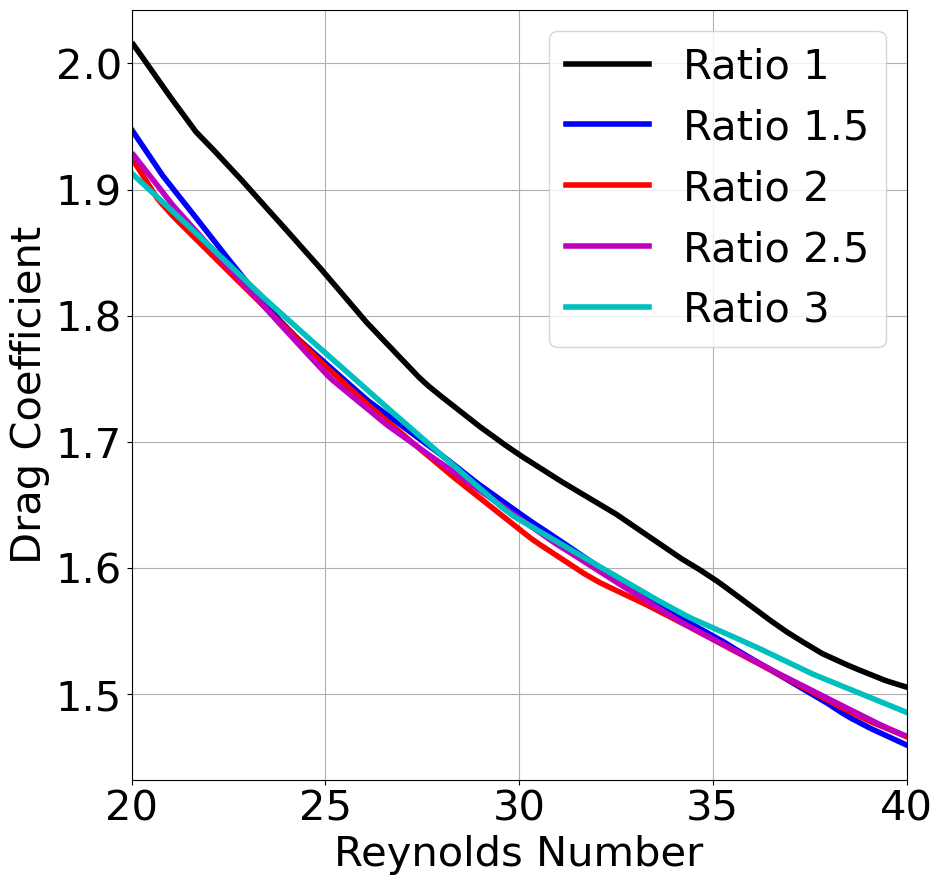}
		\caption{Angle of Attack: $120^0$}
	\end{subfigure}
	\begin{subfigure}[t]{0.32\textwidth}
		\includegraphics[width=\textwidth]{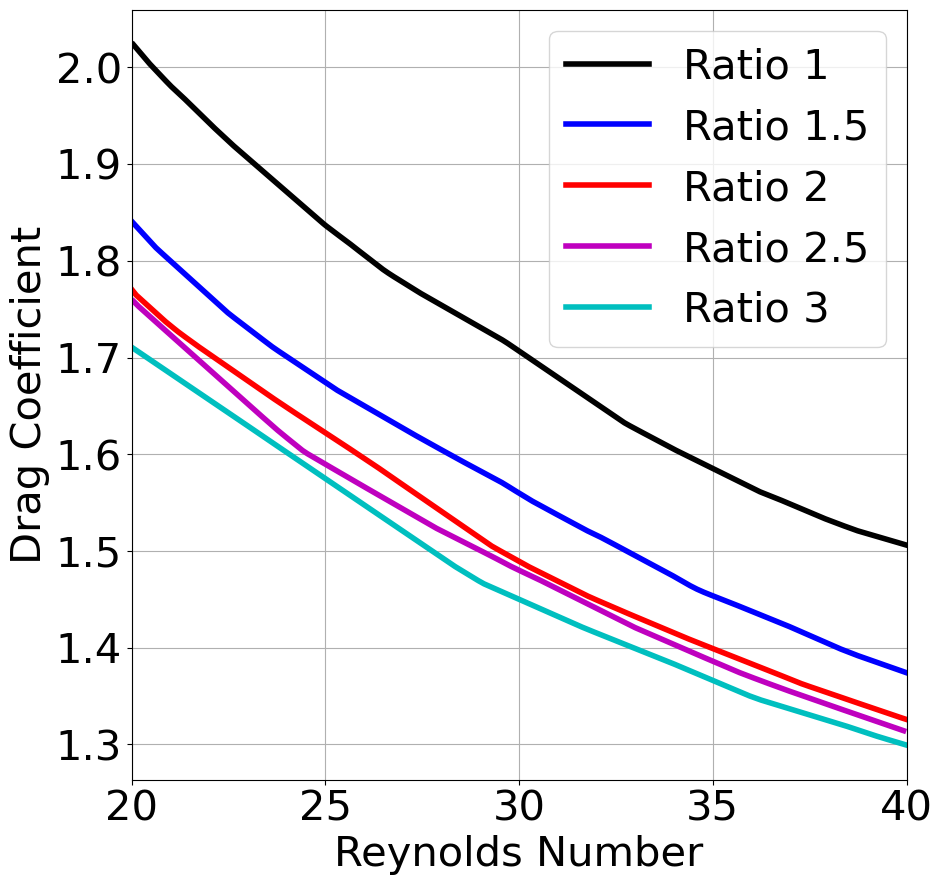}
		\caption{Angle of Attack: $135^0$}
	\end{subfigure}
	\begin{subfigure}[t]{0.32\textwidth}
		\includegraphics[width=\textwidth]{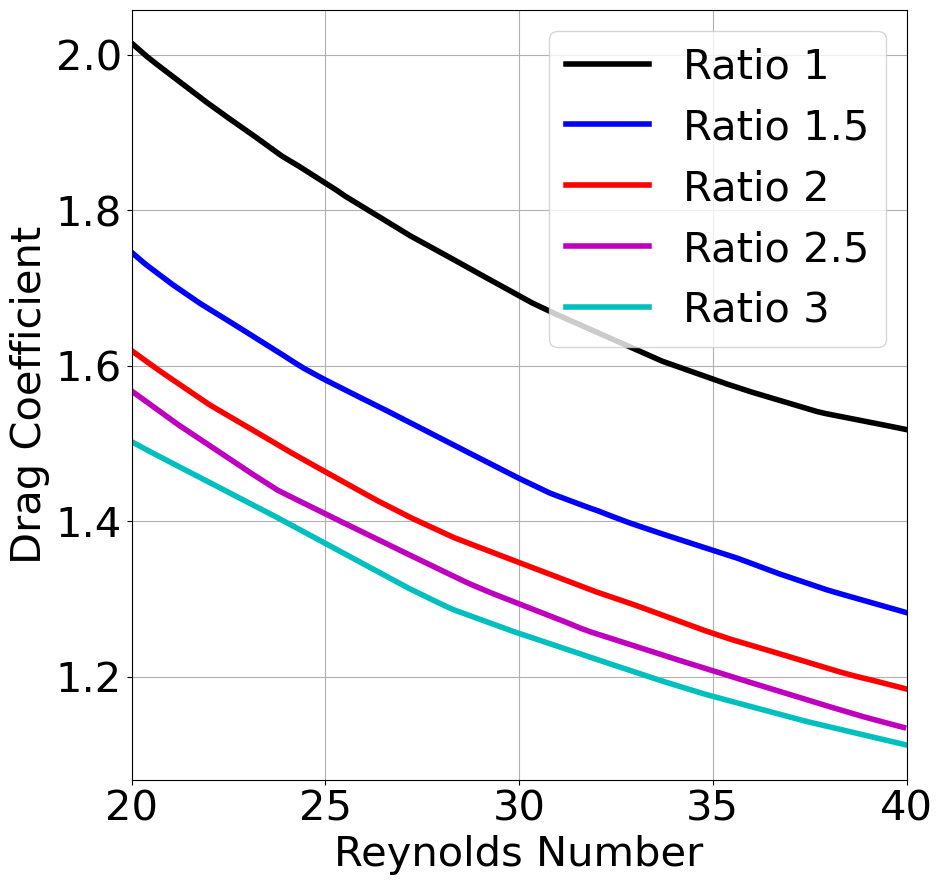}
		\caption{Angle of Attack: $150^0$}
	\end{subfigure}
	\begin{subfigure}[t]{0.32\textwidth}
		\includegraphics[width=\textwidth]{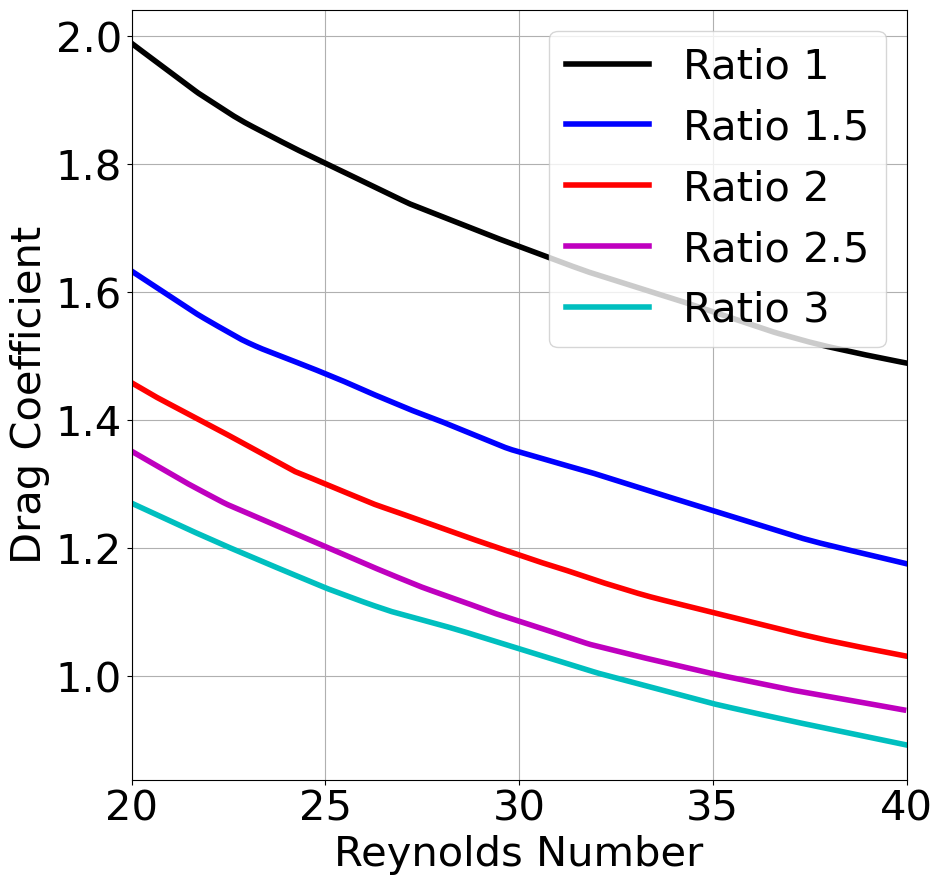}
		\caption{Angle of Attack: $180^0$}
	\end{subfigure}
	\caption{Case of Single Elliptic Cylinder: Drag Coefficient}
	\label{Fig:single ellipse drag vs reynolds varying angle}
\end{figure}
%\begin{figure}[H]
%	\centering
%	\begin{subfigure}[t]{0.32\textwidth}
%		\includegraphics[width=\textwidth]{Figures/single_ellipse/lift_drag/drag_vs_Re_ratio_1.png}
%		\caption{Ratio: Major/Minor Axis: 1}
%	\end{subfigure}
%	\begin{subfigure}[t]{0.32\textwidth}
%		\includegraphics[width=\textwidth]{Figures/single_ellipse/lift_drag/drag_vs_Re_ratio_1.5.png}
%		\caption{Ratio: Major/Minor Axis: 1.5}
%	\end{subfigure}
%	\begin{subfigure}[t]{0.32\textwidth}
%		\includegraphics[width=\textwidth]{Figures/single_ellipse/lift_drag/drag_vs_Re_ratio_2.png}
%		\caption{Ratio: Major/Minor Axis: 2}
%	\end{subfigure}
%	\begin{subfigure}[t]{0.32\textwidth}
%		\includegraphics[width=\textwidth]{Figures/single_ellipse/lift_drag/drag_vs_Re_ratio_2.5.png}
%		\caption{Ratio: Major/Minor Axis: 2.5}
%	\end{subfigure}
%	\begin{subfigure}[t]{0.32\textwidth}
%		\includegraphics[width=\textwidth]{Figures/single_ellipse/lift_drag/drag_vs_Re_ratio_3.png}
%		\caption{Ratio: Major/ Minor Axis: 3}
%	\end{subfigure}
%	\caption{Case of Single Elliptic Cylinder: Drag Coefficient}
%	\label{Fig:single ellipse drag vs reynolds varying ratio}
%\end{figure}
The effect of Reynolds number on the drag coefficient, for various angles of attack, is shown in \cref{Fig:single ellipse drag vs reynolds varying angle} for different aspect ratios. First, for the case of unity aspect ratio, there is no effect of the angle of attack since it is a circular cylinder. The drag coefficient is the same at all angles of attack and reduces with Reynolds number. For non-unity aspect ratios, interesting trends are observed. For a vertical alignment ($90^{\text{o}}$), the drag monotonically increases with aspect ratios. However, as the cylinder is rotated counterclockwise to higher angles, the curves cross those of the circular cylinder and predict a lower drag coefficient. at an angle of attack of $120^{\text{o}}$, we see that the drag for all aspect ratios higher than unity is almost similar. When the cylinder reaches a horizontal position ($180^{\text{o}}$), the drag becomes significantly smaller than the circular cylinder due to streamlined body. As the aspect ratio is increased, the cylinder tends to get slimmer, so the drag due to wake formation and skin friction is reduced.
\par For tandem cylinders, there are six parameters, and hence the lift and drag coefficients vary over a six-dimensional space. It is difficult to analyze the variations in a six dimensional input space due to a complex interactions between these inputs. Hence, we have arbitrarily selected two cases to demonstrate what can be obtained from the trained neural networks. It can be seen that some variations seem to be linear whereas others show a significantly nonlinear behavior. We also observe that the coefficients on the downstream cylinder are substantially different than the upstream cylinder which shows that the upstream cylinder perturbs the uniform flow field. We are currently developing a graphical interface which can be used to maneuver the complete space and easily compute the parametric variations. This interface can be used as a virtual laboratory experiment in fluid mechanics education.

\begin{figure}[H]
	\centering
	\begin{subfigure}[t]{0.32\textwidth}
		\includegraphics[width=\textwidth]{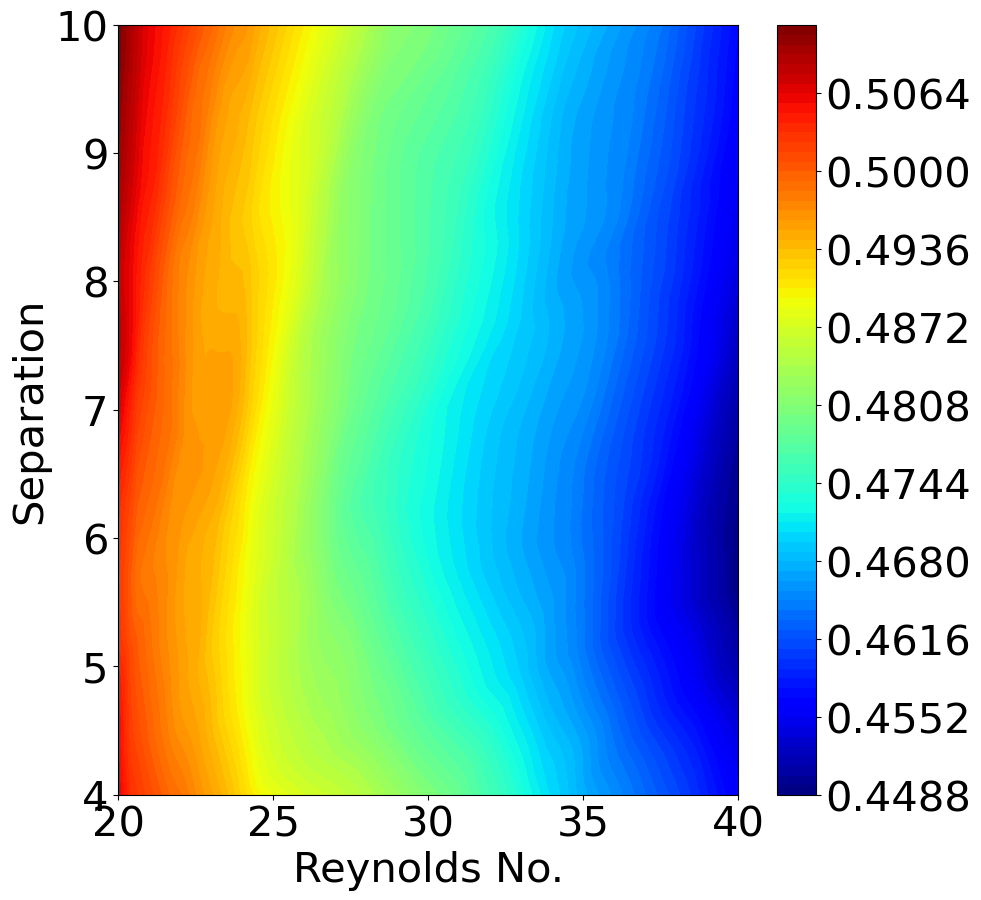}
		\caption{Upstream and Downstream Angles: $135^0$}
	\end{subfigure}
	\begin{subfigure}[t]{0.32\textwidth}
		\includegraphics[width=\textwidth]{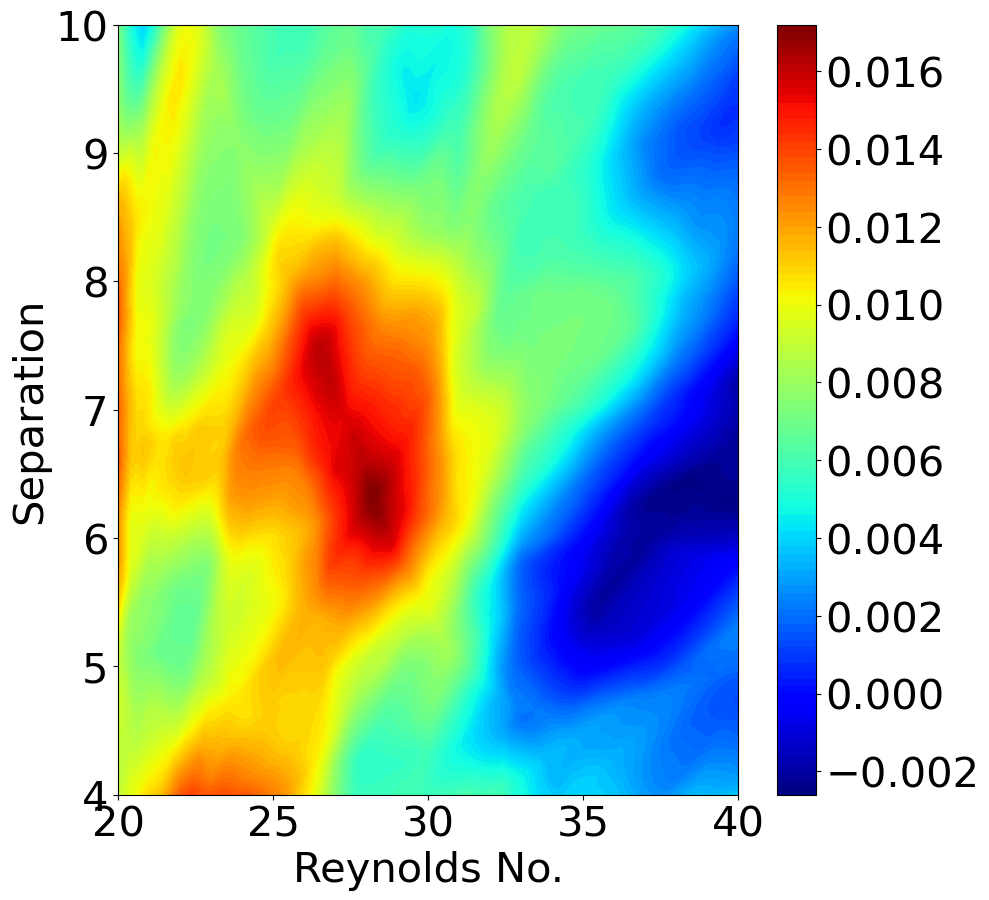}
		\caption{Upstream Angle: $90^0$, Downstream Angle: $135^0$}
	\end{subfigure}
	\begin{subfigure}[t]{0.32\textwidth}
		\includegraphics[width=\textwidth]{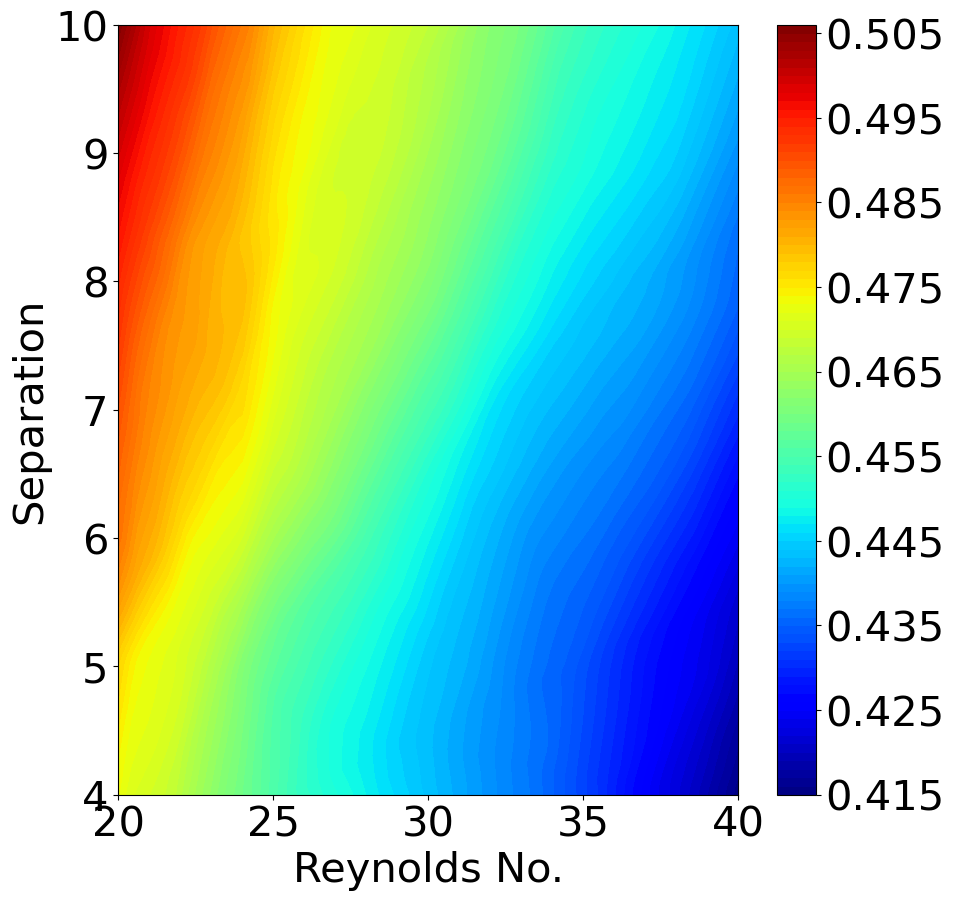}
		\caption{Upstream Angle: $135^0$, Downstream Angle: $90^0$}
	\end{subfigure}
	\caption{Case of Double Elliptic Cylinder: Lift Coefficient of Upstream Cylinder (Upstream and Downstream Aspect Ratio: 2)}
	\label{Fig:double ellipse lift upstream vs Re_separation}
\end{figure}

\begin{figure}[H]
	\centering
	\begin{subfigure}[t]{0.32\textwidth}
		\includegraphics[width=\textwidth]{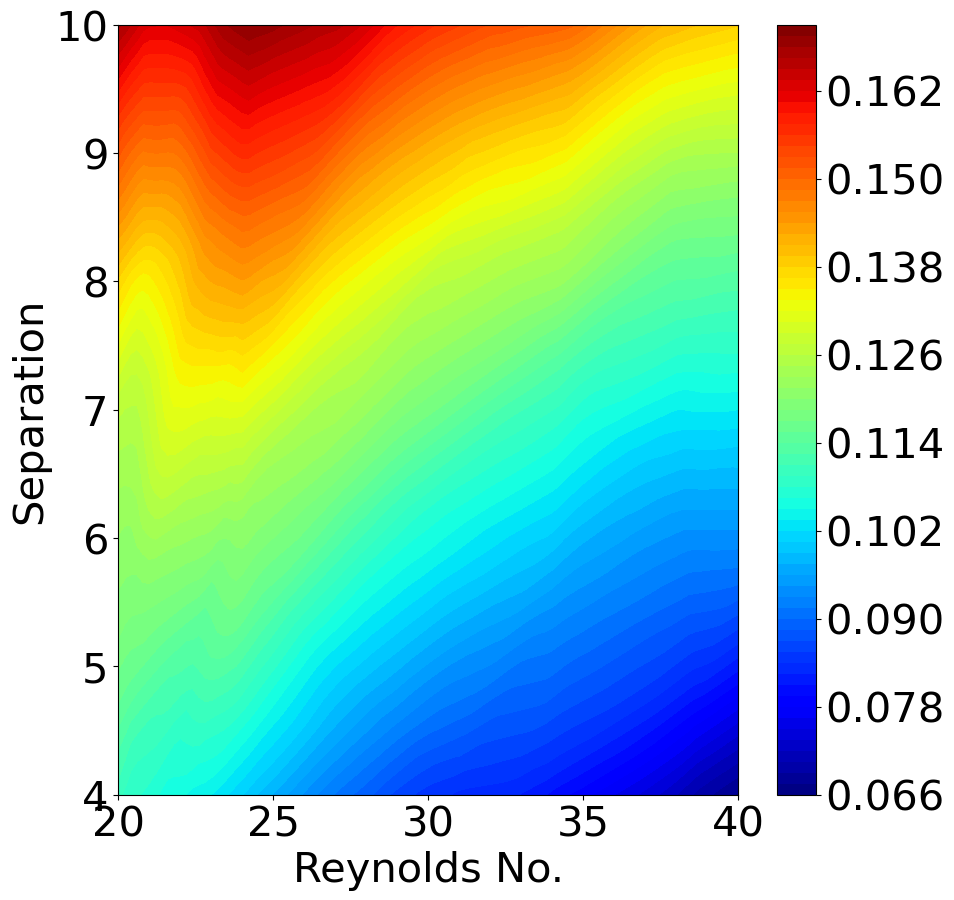}
		\caption{Upstream and Downstream Angles: $135^0$}
	\end{subfigure}
	\begin{subfigure}[t]{0.32\textwidth}
		\includegraphics[width=\textwidth]{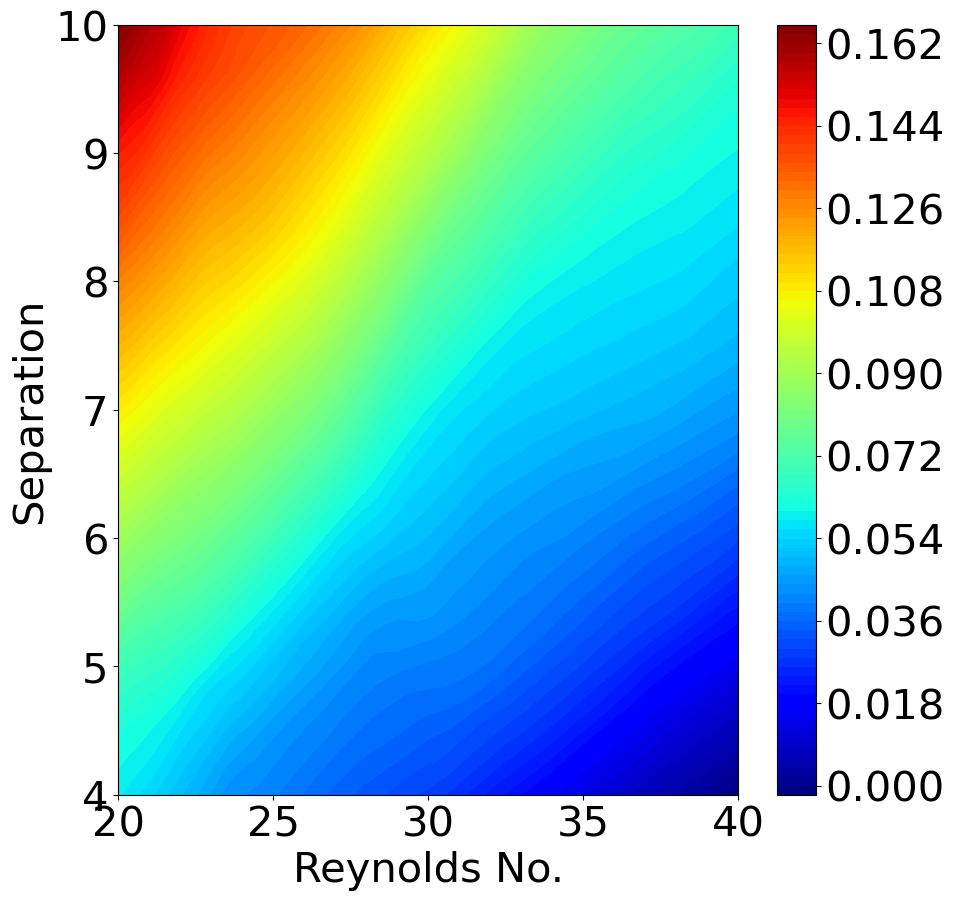}
		\caption{Upstream Angle: $90^0$, Downstream Angle: $135^0$}
	\end{subfigure}
	\begin{subfigure}[t]{0.32\textwidth}
		\includegraphics[width=\textwidth]{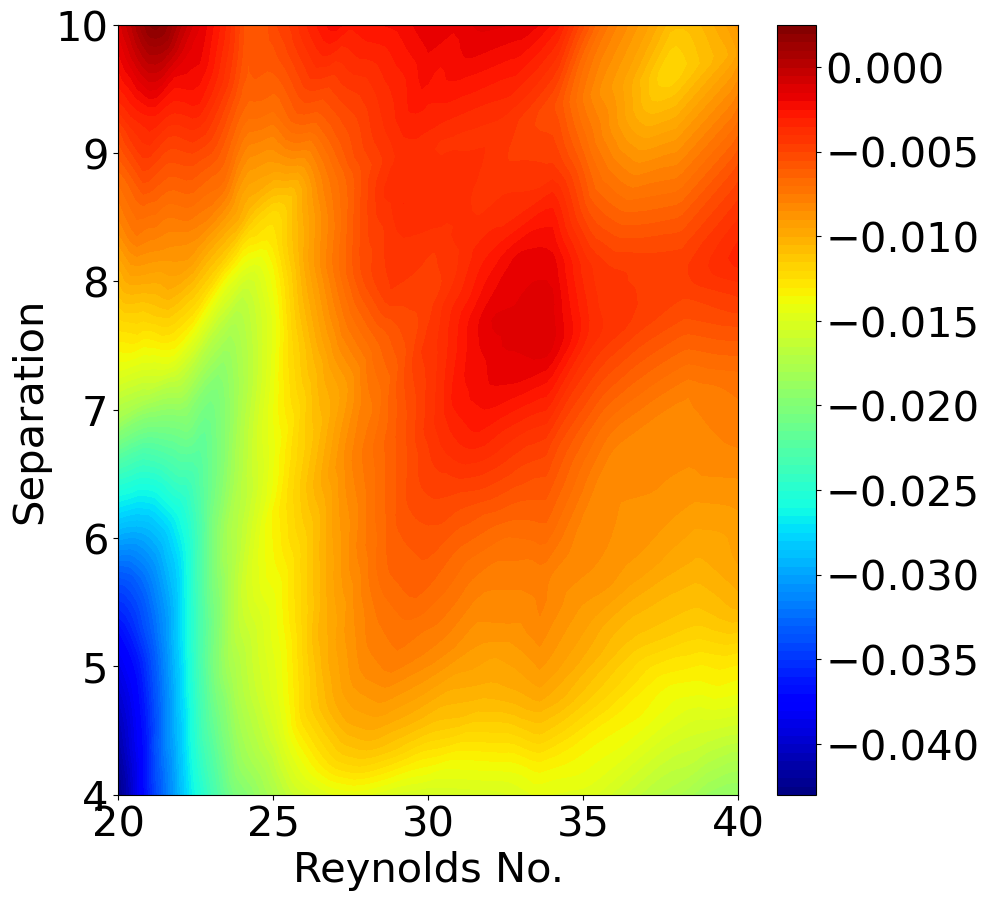}
		\caption{Upstream Angle: $135^0$, Downstream Angle: $90^0$}
	\end{subfigure}
	\caption{Case of Double Elliptic Cylinder: Lift Coefficient of Downstream Cylinder (Upstream and Downstream Aspect Ratio: 2)}
	\label{Fig:double ellipse lift downstream vs Re_separation}
\end{figure}

\begin{figure}[H]
	\centering
	\begin{subfigure}[t]{0.32\textwidth}
		\includegraphics[width=\textwidth]{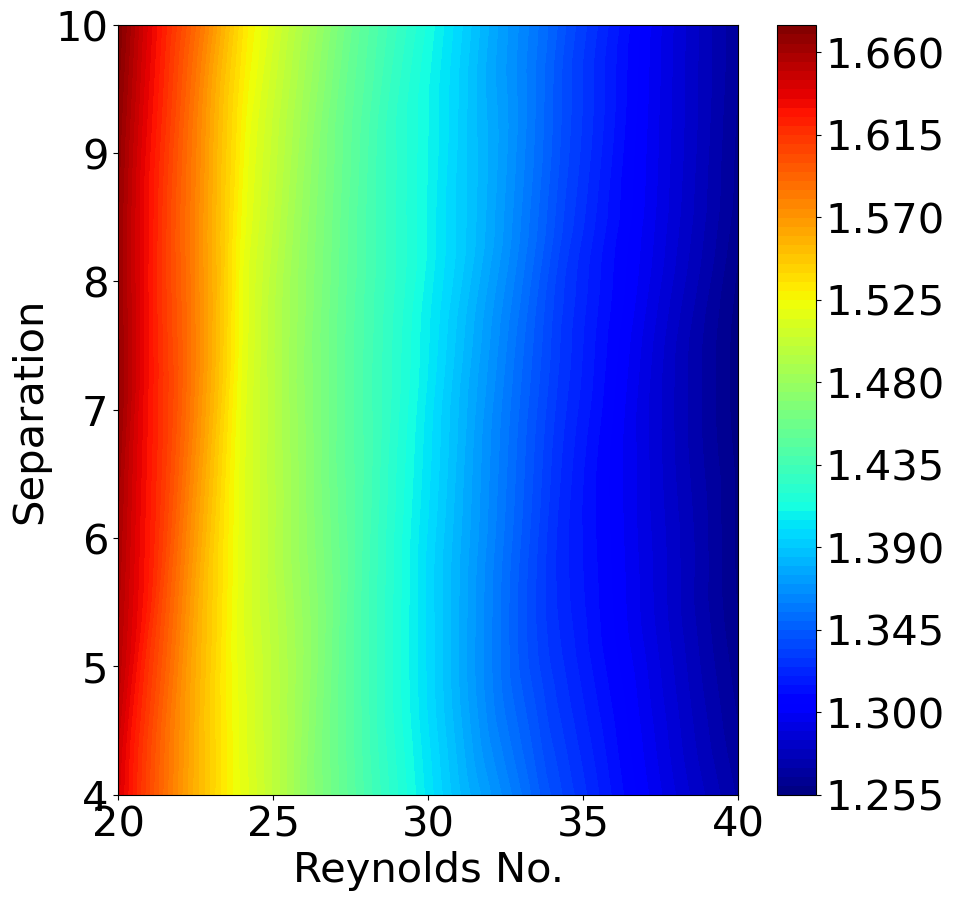}
		\caption{Upstream and Downstream Angles: $135^0$}
	\end{subfigure}
	\begin{subfigure}[t]{0.32\textwidth}
		\includegraphics[width=\textwidth]{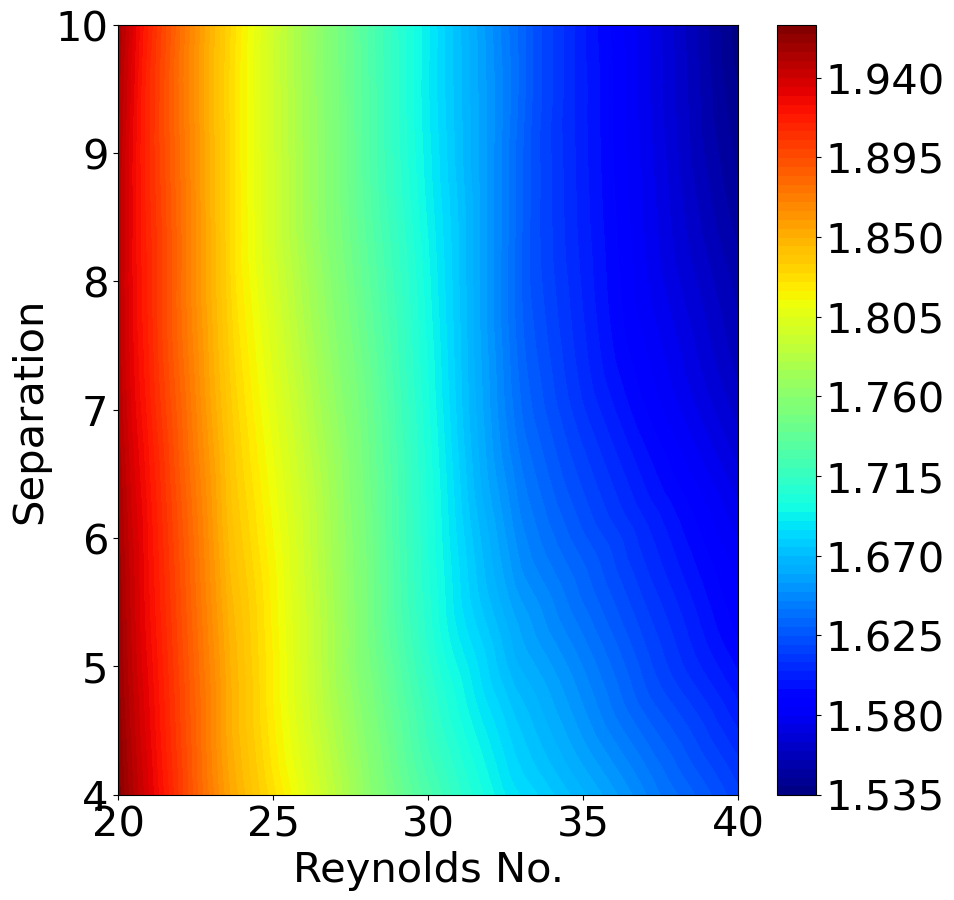}
		\caption{Upstream Angle: $90^0$, Downstream Angle: $135^0$}
	\end{subfigure}
	\begin{subfigure}[t]{0.32\textwidth}
		\includegraphics[width=\textwidth]{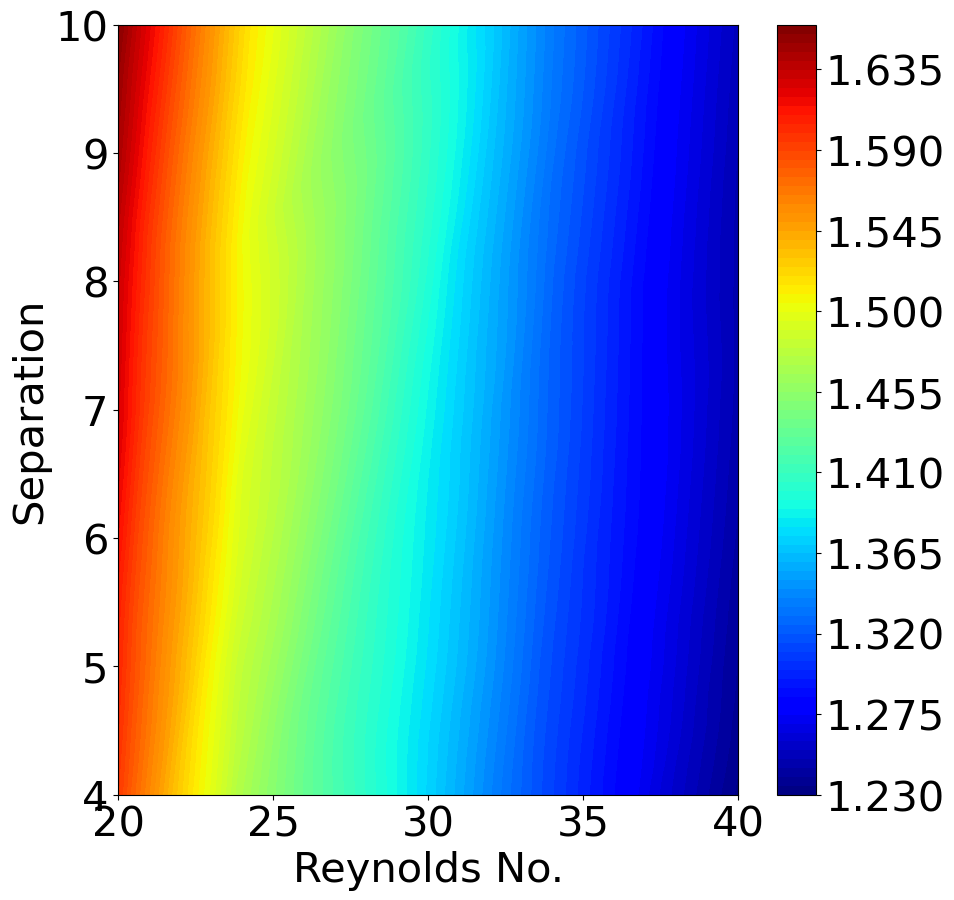}
		\caption{Upstream Angle: $135^0$, Downstream Angle: $90^0$}
	\end{subfigure}
	\caption{Case of Double Elliptic Cylinder: Drag Coefficient of Upstream Cylinder (Upstream and Downstream Aspect Ratio: 2)}
	\label{Fig:double ellipse drag upstream vs Re_separation}
\end{figure}

\begin{figure}[H]
	\centering
	\begin{subfigure}[t]{0.32\textwidth}
		\includegraphics[width=\textwidth]{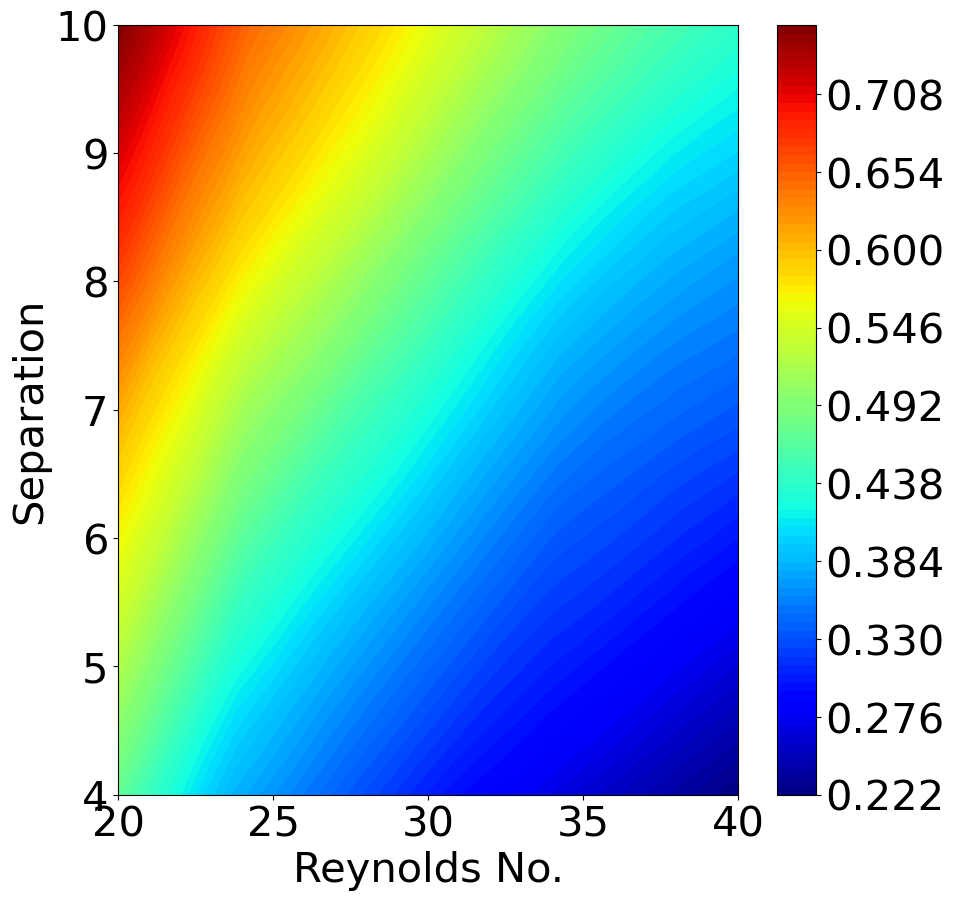}
		\caption{Upstream and Downstream Angles: $135^0$}
	\end{subfigure}
	\begin{subfigure}[t]{0.32\textwidth}
		\includegraphics[width=\textwidth]{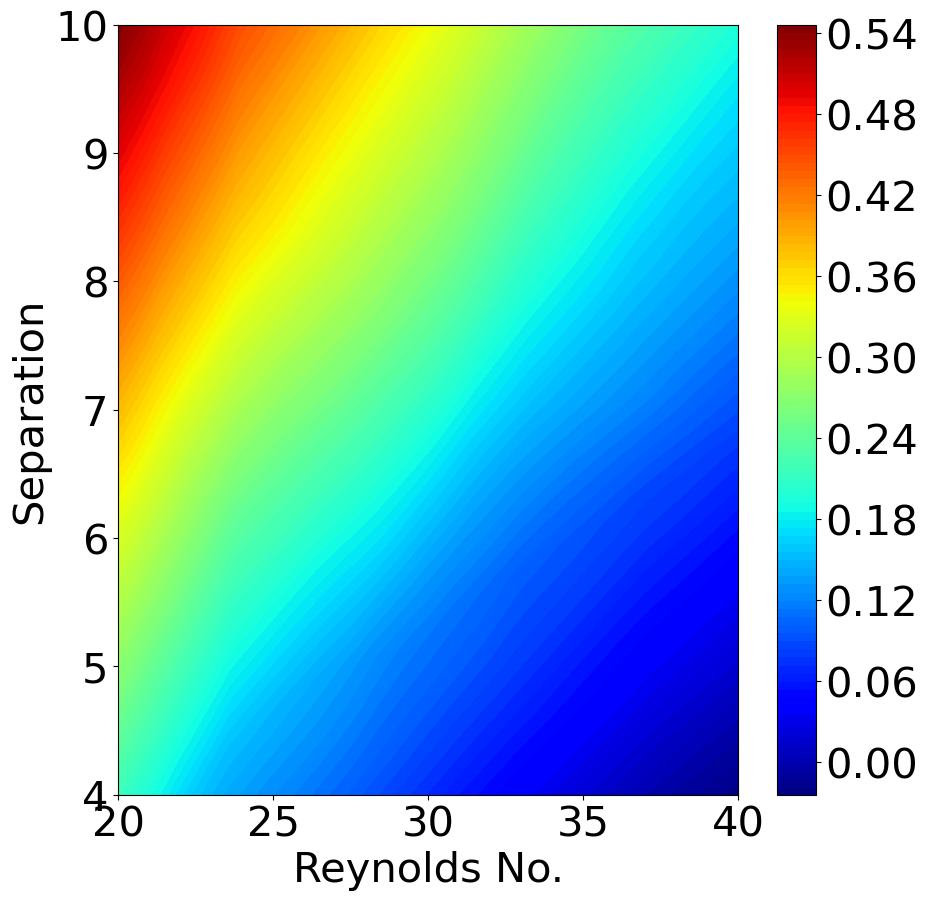}
		\caption{Upstream Angle: $90^0$, Downstream Angle: $135^0$}
	\end{subfigure}
	\begin{subfigure}[t]{0.32\textwidth}
		\includegraphics[width=\textwidth]{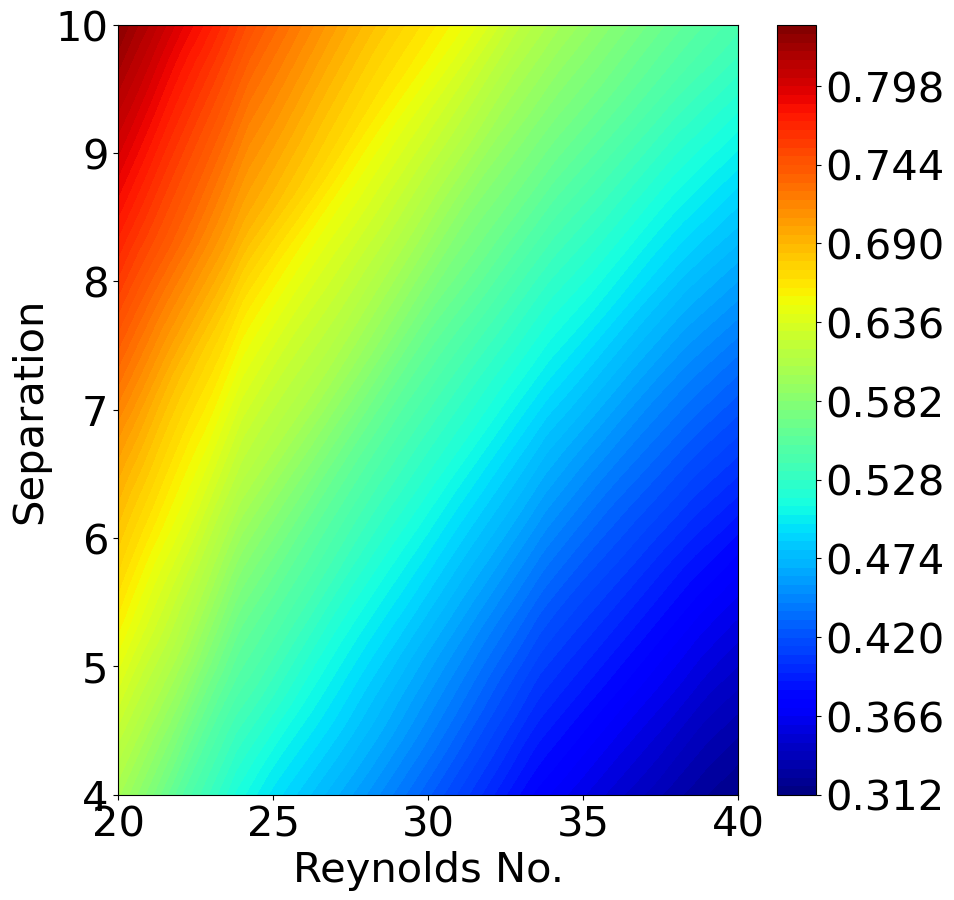}
		\caption{Upstream Angle: $135^0$, Downstream Angle: $90^0$}
	\end{subfigure}
	\caption{Case of Double Elliptic Cylinder: Drag Coefficient of Downstream Cylinder (Upstream and Downstream Aspect Ratio: 2)}
	\label{Fig:double ellipse drag downstream vs Re_separation}
\end{figure}
\section{Sensitivity Analysis}\label{Sec:Sensitivity Analysis}
Sensitivity can be defined as the effect of perturbation of an input on an output. Let $Y$ be a scalar output dependent on a $d$ dimensional input vector $\bm{X} = [X_1,X_2,\dots X_d]$. Let $f$ denote the model which relates $\bm{X}$ to $Y$ i.e., $Y=f(\bm{X})$. For instance, in the case of single elliptic cylinder, drag coefficient is a particular output ($Y$) as a function of inputs such as Reynolds number, ratio and angle of attack (3 dimensional vector $\bm{X}$) and the neural network is used as the model ($f$). Partial derivative of $Y$ with respect to a particular input $X_i$ is one way to define the sensitivity of $Y$ to $X_i$. Since the partial derivative has to be evaluated at a particular value of $\bm{X}=\bm{\hat{X}}$, this gives an estimate of the local sensitivity at $\bm{\hat{X}}$. Response surfaces can be used to visualize local sensitivity. For example, slope of the contour lines in the \cref{Fig:single ellipse drag vs ratio_angle Re 20} is low for ratio of 1.25 and $0^\text{o}$ angle of attack. Hence, in this region, the drag is more sensitive to the ratio than angle of attack. On the other hand, in the region of 2.5 ratio and $45^\text{o}$ angle of attack, the contours are steep indicating that the drag is more sensitive to the angle. This example demonstrates that the partial derivatives can vary significantly from one design point to another due to the nonlinearity of the model. Hence, for practical problems with nonlinear relationships, the local sensitivity analysis does not give any information of the entire design space.
\par
\begin{figure}[H]
	\centering
	\includegraphics[width=0.4\textwidth]{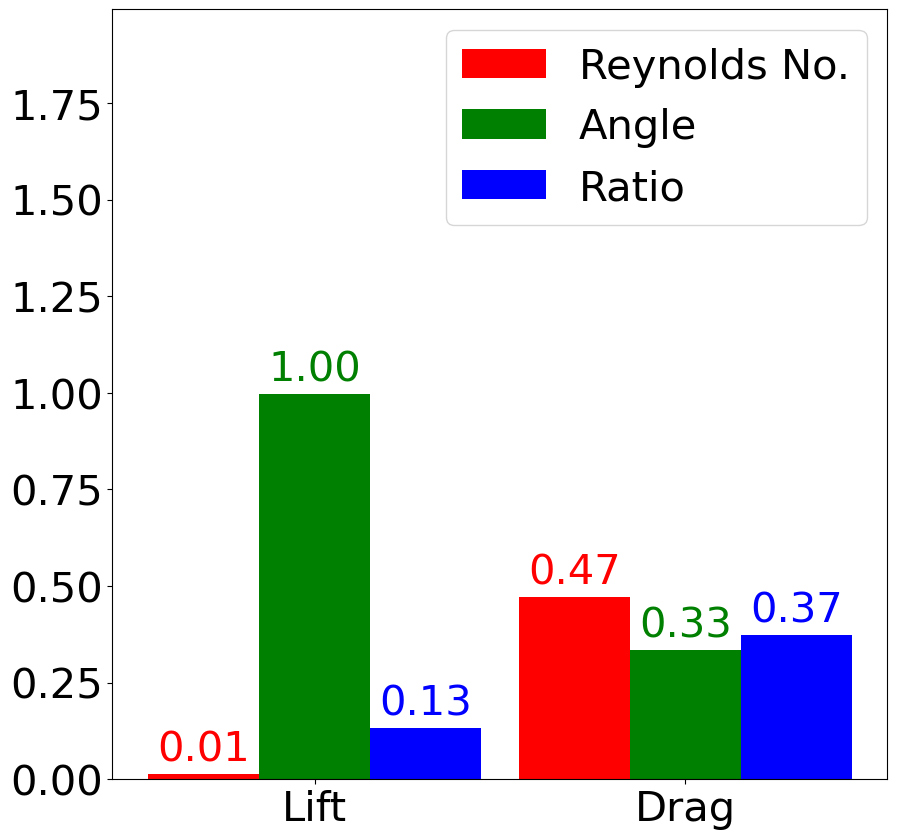}
	\caption{Case of Single Elliptic Cylinder: Total Sobol Indices: Sensitivity of Lift and Drag to 3 Inputs}
	\label{Fig:single ellipse Total Sobol Indices}
\end{figure}

 In this work, we use the variance based analysis, also known as Sobol method \cite{sobol2001global} to estimate the global sensitivities of each output with respect to each input. The relation $Y=f(\bm{X})$ can be written as a summation of functions over individual inputs:
\begin{equation}
	Y = f(\bm{X}) = f_0 + \sum_{i=1}^{d} f_i(X_i) + \sum_{i<j}^{d} f_{i,j}(X_i,X_j) + \dots + f_{1,2,\dots,d}(X_1,X_2,\dots X_d)
	\label{Eq:sobol decomp}
\end{equation}
where, $f_0$ is a constant, $f_i$ is a function of single input $X_i$, $f_{i,j}$ is a function of two inputs $X_i$ and $X_j$ and so on. This summation has $2^d$ functions for a $d$ dimensional input space. The decomposition is known as ANOVA (analysis of variances) if each of the functions has zero means:
\begin{equation}
	\int f_{i_1,i_2,\dots,i_s} (X_{i_1}, X_{i_2}, \dots, X_{i_s}) dX_k = 0 \hspace{0.2cm} \text{for} \hspace{0.2cm} k=i_1,i_2,\dots,i_s
\end{equation}
It can be shown that if the above condition is satisfied, the functions are orthogonal and thus, the decomposition is unique \cite{saltelli2008global}. If $f(\bm{X})$ is assumed to be square-integrable, squaring \cref{Eq:sobol decomp} and integrating gives:
\begin{equation}
	\int Y^2 d\bm{X}  - f_0^2 = \sum_{s=1}^{d} \sum_{i_1<\dots<i_s}^{d} \int f^2_{i_1,\dots,i_s} d X_{i_1} \dots X_{i_s}
	\label{Eq:sobol decomp integrate}
\end{equation}
Note that the cross terms such as $\int f_{i_1}f_{i_2} dX_{i_1}dX_{i_2}$ are zero due to orthogonality and only the squared terms remain. The left hand side of \cref{Eq:sobol decomp integrate} is equal to variance of $Y$ and the right hand side is a summation of variances due to groups of inputs. Hence, the total variance in $Y$ is decomposed into variances attributed to individual inputs and interactions between them:
\begin{equation}
	Var(Y) = \sum_{i=1}^{d} V_i + \sum_{i<j}^{d} V_{i,j} + \dots + V_{1,2,\dots,d}
	\label{Eq:sobol decomp variance}
\end{equation}
First order sensitivity index is defined as $S_i=V_i/Var(Y)$. Higher order indices such as $S_{i,j}$ are similarly defined. From \cref{Eq:sobol decomp variance}, it can be seen that these ($2^d-1$) indices are non-negative and sum to unity. Total Sobol index ($S_{T_i}$) for each input $X_i$ is defined as sum of all the first and higher order indices with $X_i$ in it. For example, for the case with 3 inputs, $S_{T_1} = S_1 + S_{1,2} + S_{1,3} + S_{1,2,3}$. Note that the sum of total indices is typically more than unity since the interaction terms are counted more than once. In this work, we analyze the total Sobol indices of each output (lift and drag coefficients) with respect to each input (Reynolds number, angle, aspect ratio and separation). For simple functions, the integrals in \cref{Eq:sobol decomp integrate} can be evaluated analytically. For practical cases however, Monte-Carlo methods are used to estimate total Sobol indices. Brute force computation is $\mathcal O (N^2)$ where, $N$ denotes the number of Monte-Carlo samples. Since $N$ can of the order of $10^5\sim10^6$, these computations are quite expensive even when surrogate models are used. Hence, we use the algorithm proposed by \citet{saltelli2008global} which requires $\mathcal O (N(d+2))$ computations.
\begin{figure}[H]
	\centering
	\begin{subfigure}[t]{0.49\textwidth}
		\includegraphics[width=\textwidth]{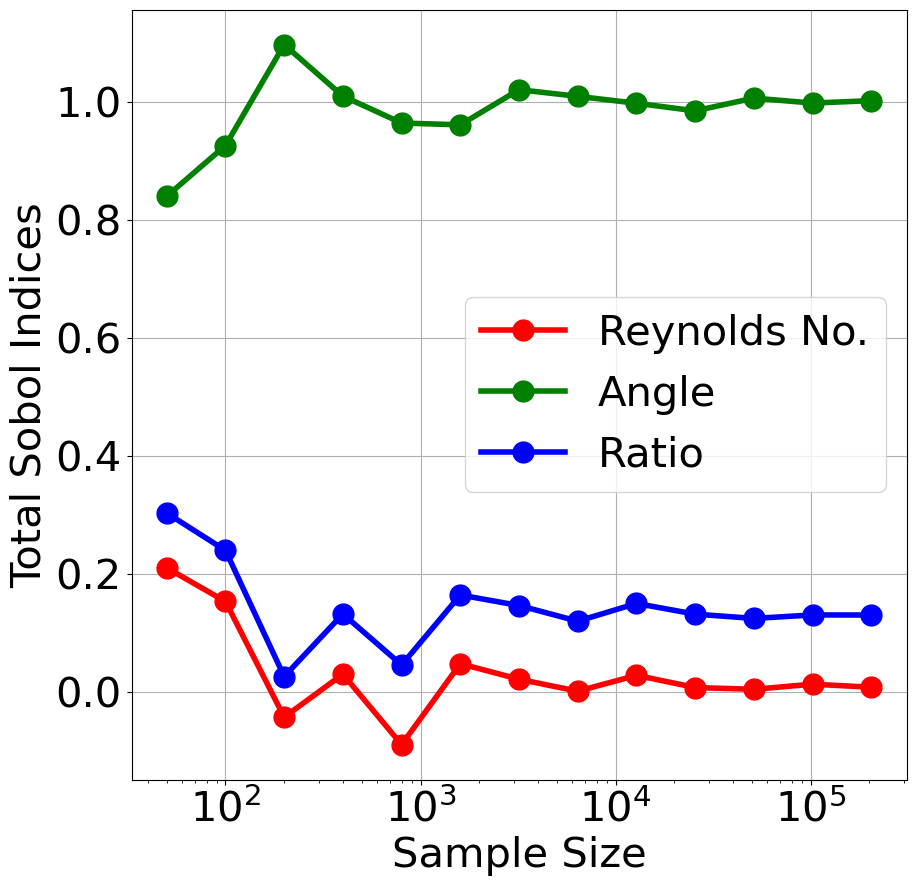}
		\caption{Lift Coefficient}
	\end{subfigure}
	\begin{subfigure}[t]{0.49\textwidth}
		\includegraphics[width=\textwidth]{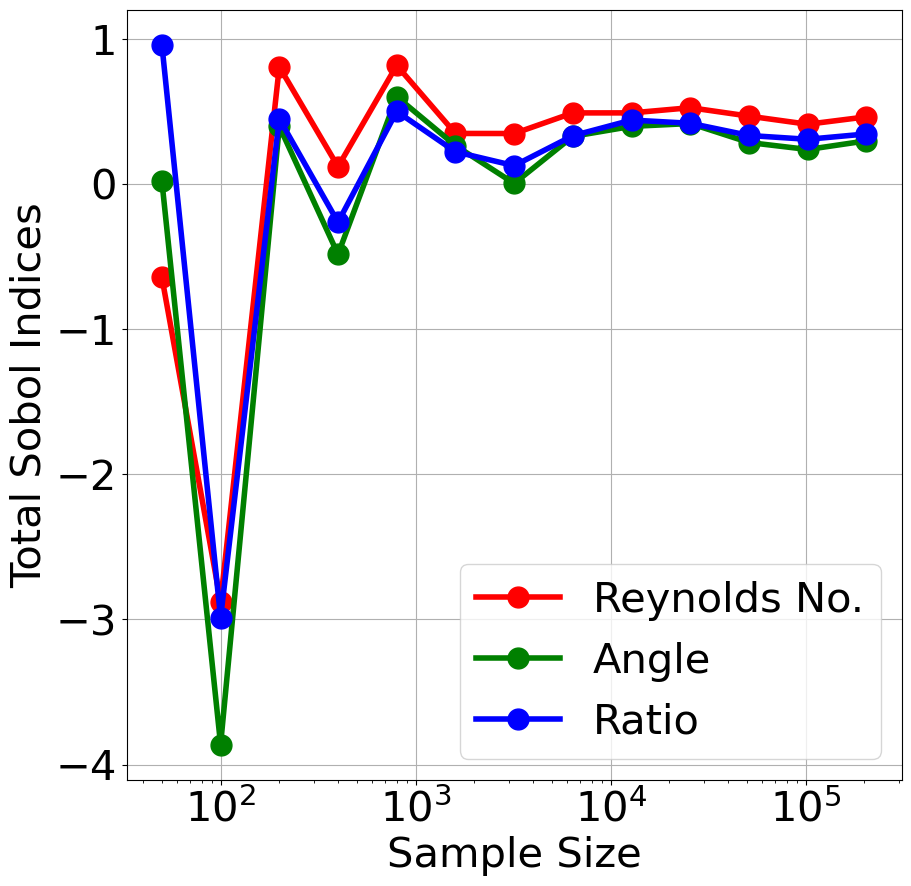}
		\caption{Drag Coefficient}
	\end{subfigure}
	\caption{Case of Single Elliptic Cylinder: Estimation of Total Sobol Indices: Convergence of Monte-Carlo Method}
	\label{Fig:single ellipse Estimation of Total Sobol Indices: Convergence of Monte-Carlo Method}
\end{figure}
The Monte-Carlo method obtains numerical estimates by repeated random sampling. The convergence error is $\mathcal{O} (1/\sqrt{N})$ \cite{caflisch1998monte} where, $N$ is the sample size. In the absence of analytical solution, the sample size is increased until the estimates converge asymptotically. For the case of single cylinder, there are two outputs (lift and drag coefficients) and three inputs (Reynolds number, angle, aspect ratio). Thus, convergence of $3\times2=6$ total Sobol indices is plotted in \cref{Fig:single ellipse Estimation of Total Sobol Indices: Convergence of Monte-Carlo Method}. The sample size is increased exponentially from 50 to 2E5 by a factor of 2 each time. The lift and drag coefficients are estimated as a function of randomly generated sets of inputs using the MLPNN described before. The computational efficiency of neural network facilitates such high sample sizes. This demonstrates the benefit of coupling neural networks with numerical simulations. The initial estimates are inaccurate (sometimes negative) but stationarity is obtained as the sample size reaches close to 1E4. Thus, the average of last 5 estimates is recorded as the converged value. \Cref{Fig:single ellipse Total Sobol Indices} is a bar chart with sensitivity of each output with respect to each input. As discussed before, the sum of total Sobol indices of both outputs is greater than unity. It can be seen that the lift coefficient is highly sensitive to the angle of attack and its dependence on the Reynolds number and aspect ratio is negligible. On the other hand, the drag is more sensitive to the Reynolds number but its sensitivity to the other inputs is also important.

\begin{figure}[H]
	\centering
	\begin{subfigure}[t]{0.49\textwidth}
		\includegraphics[width=\textwidth]{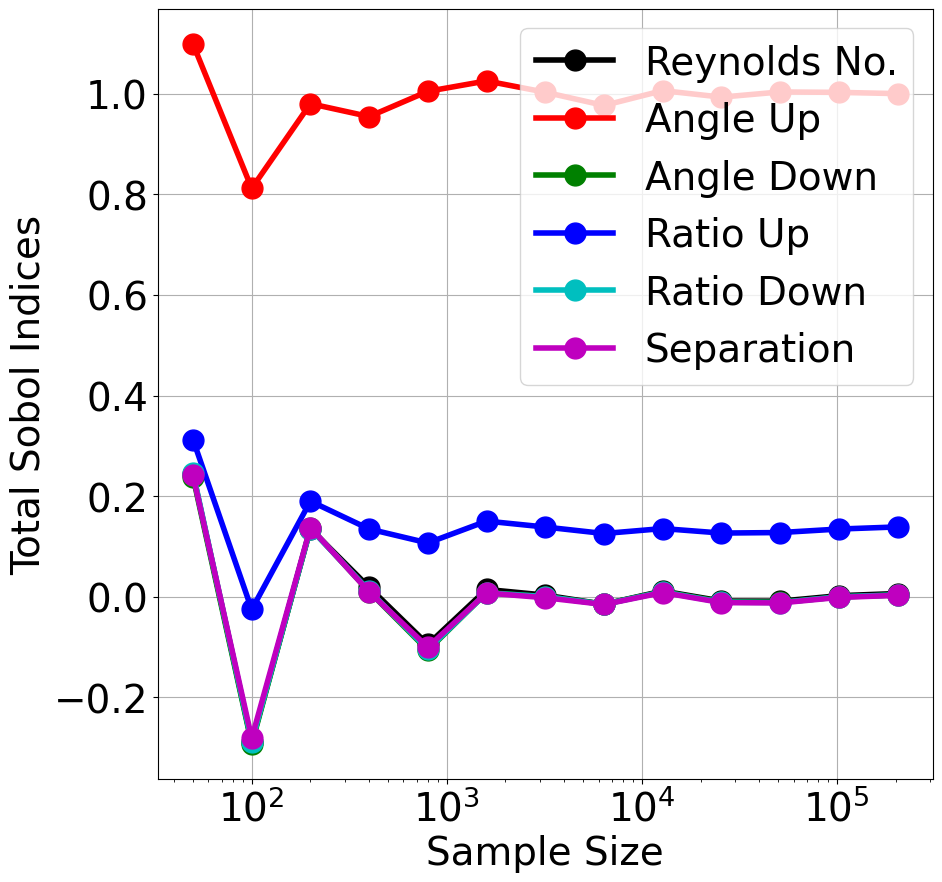}
		\caption{Upstream Lift Coefficient}
	\end{subfigure}
	\begin{subfigure}[t]{0.49\textwidth}
		\includegraphics[width=\textwidth]{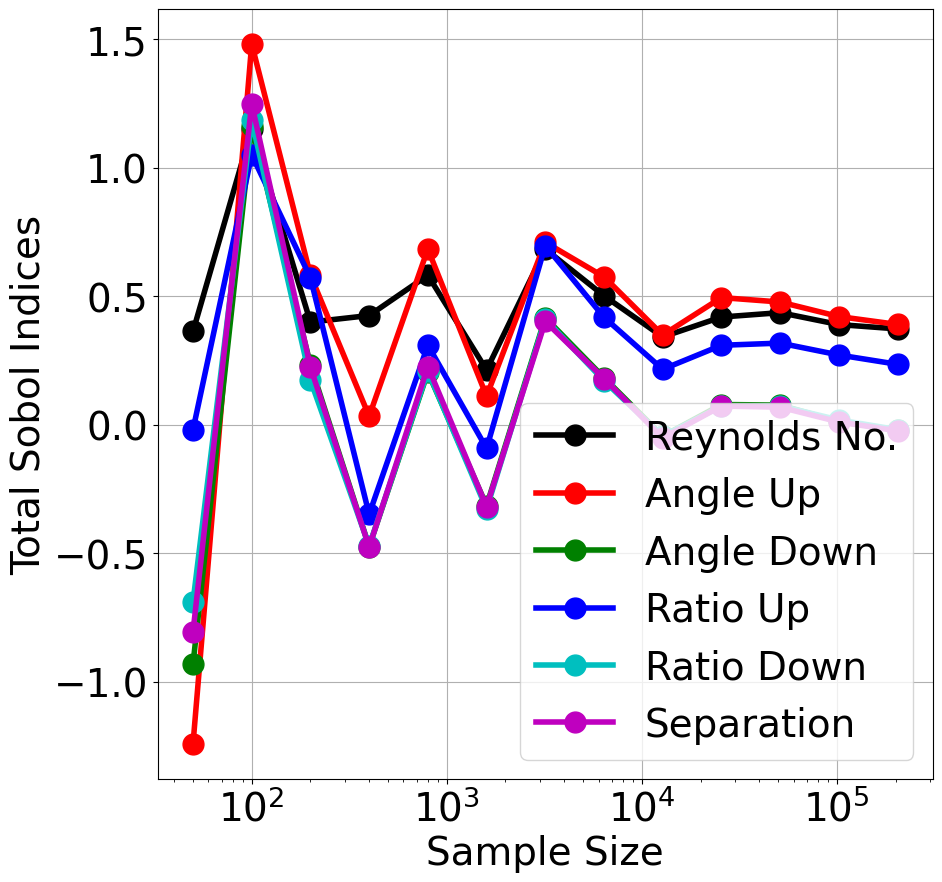}
		\caption{Upstream Drag Coefficient}
	\end{subfigure}
	\begin{subfigure}[t]{0.49\textwidth}
		\includegraphics[width=\textwidth]{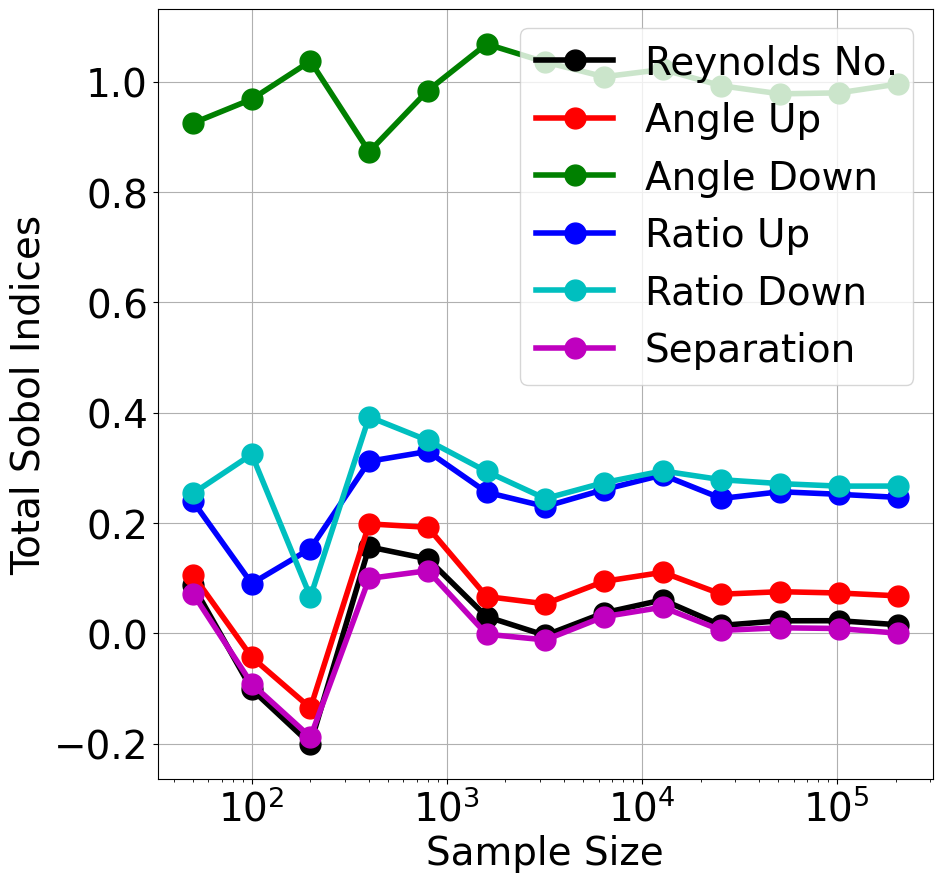}
		\caption{Downstream Lift Coefficient}
	\end{subfigure}
	\begin{subfigure}[t]{0.49\textwidth}
		\includegraphics[width=\textwidth]{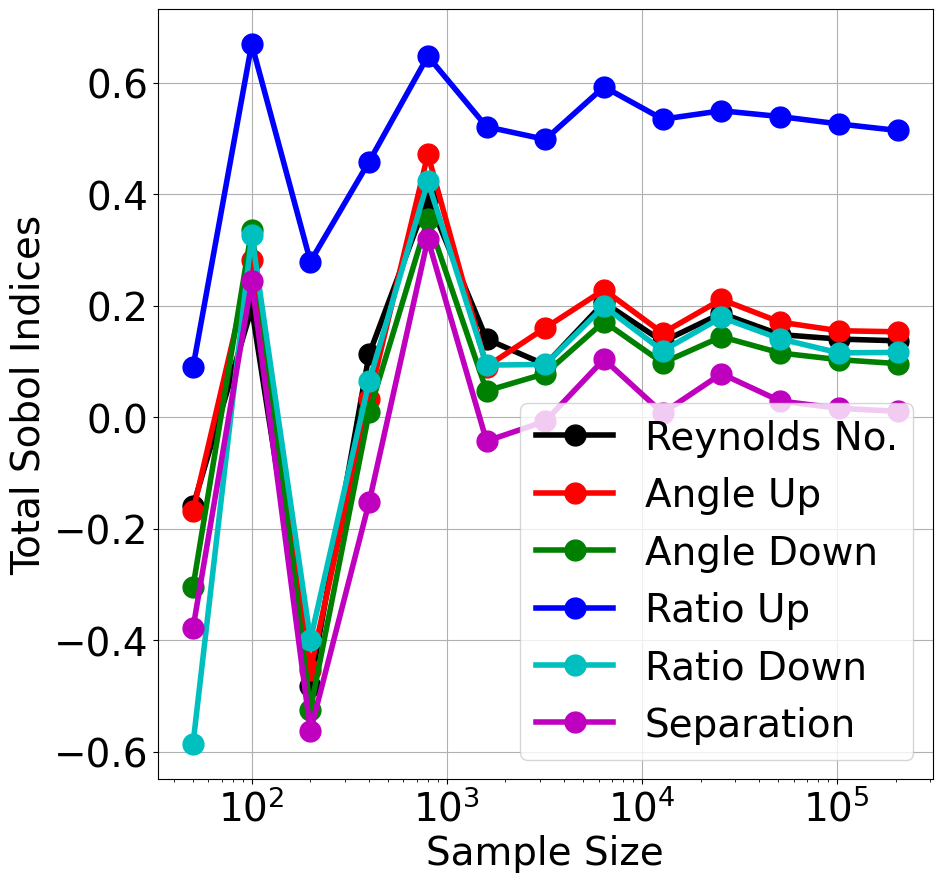}
		\caption{Downstream Drag Coefficient}
	\end{subfigure}
	\caption{Case of Double Elliptic Cylinder: Estimation of Total Sobol Indices: Convergence of Monte-Carlo Method}
	\label{Fig:double ellipse Estimation of Total Sobol Indices: Convergence of Monte-Carlo Method}
\end{figure}
\begin{figure}[H]
	\centering
	\begin{subfigure}[t]{0.49\textwidth}
		\includegraphics[width=\textwidth]{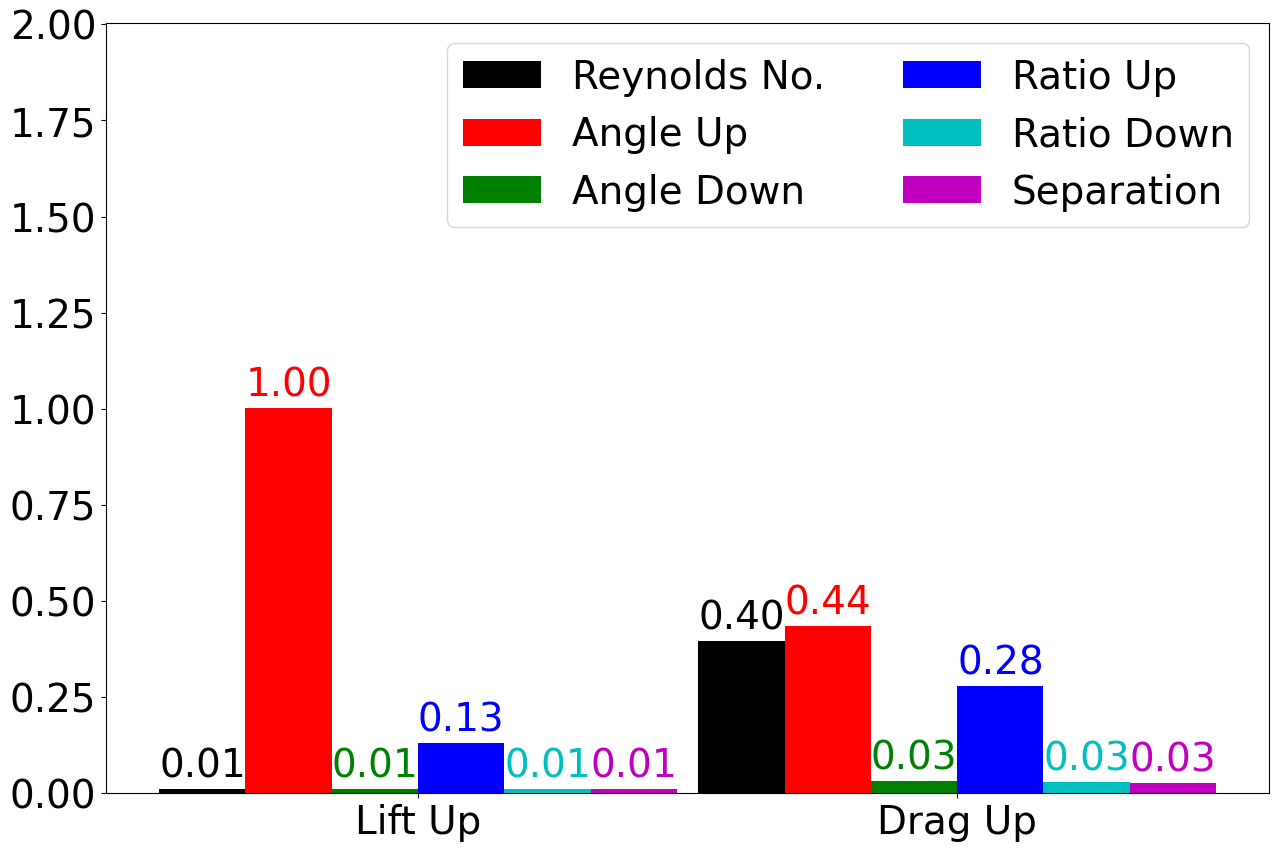}
		\caption{Upstream Lift and Drag}
		\label{Fig:double ellipse Total Sobol Indices up}
	\end{subfigure}
	\begin{subfigure}[t]{0.49\textwidth}
		\includegraphics[width=\textwidth]{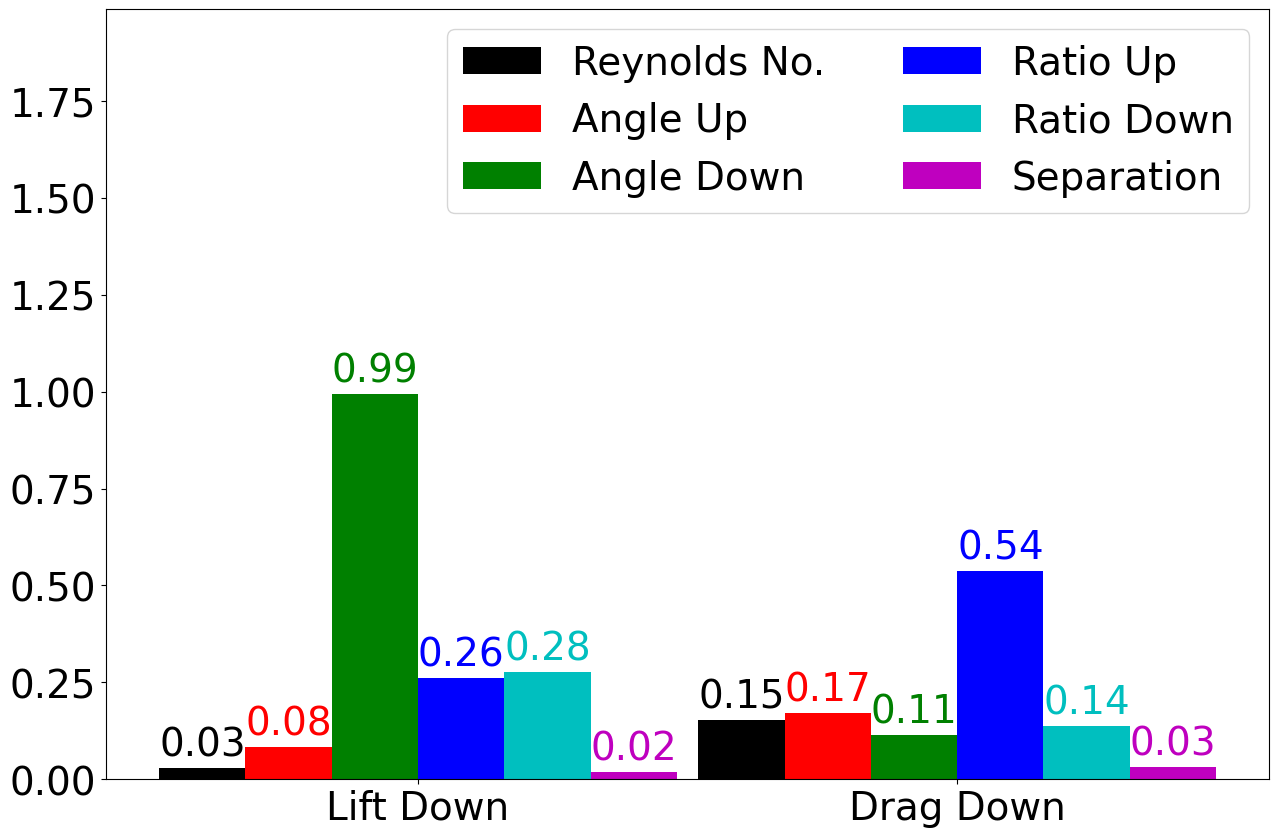}
		\caption{Downstream Lift and Drag}
		\label{Fig:double ellipse Total Sobol Indices down}
	\end{subfigure}
	\caption{Case of Double Elliptic Cylinder: Total Sobol Indices: Sensitivity of Lift and Drag to 6 Inputs}
	\label{Fig:double ellipse Total Sobol Indices}
\end{figure}

For the case of double cylinders, there are 4 outputs and 6 inputs as shown in \cref{Fig:MLPNN for Estimation of Lift and Drag Coefficients double}. Hence, there are $6\times4 = 24$ total Sobol indices. \Cref{Fig:double ellipse Estimation of Total Sobol Indices: Convergence of Monte-Carlo Method} plots the convergence for increase in sample size by a factor of 2 starting from 50. The stationarity is reached beyond 1E4 samples. Average of last five samples is recorded as the converged estimate in \cref{Fig:double ellipse Total Sobol Indices}. Similar to the case of single cylinder, the lift coefficients of both the cylinders are highly sensitive to their respective angles of attack. The lift and drag of the upstream cylinder are not sensitive to the parameters of the downstream cylinder. This shows that the downstream cylinder has negligible effect on the upstream cylinder. Moreover, these sensitivity indices in \cref{Fig:double ellipse Total Sobol Indices up} are fairly close to the indices of single cylinder in \cref{Fig:single ellipse Total Sobol Indices}. On the other hand the downstream cylinder is affected significantly by the upstream cylinder. Thus, the downstream drag is most sensitive to the upstream ratio. Hence, the Sobol indices give insight in the system and highlight the underlying physics. The input parameters with higher indices should be tightly controlled since their stochastic variation affects the output significantly. Those inputs with lower indices can be loosely controlled as their impact on the output is minimal. This information can be used practically during the design and manufacturing stages.
\section{Summary and Future Work}\label{Sec:Conclusion}
In the present paper, we have trained multilayer perceptron neural networks (MLPNN) for prediction of lift and drag coefficients due to flow over single and tandem elliptic cylinders of arbitrary aspect ratios, angles of attack, inter-cylinder distance and flow Reynolds number. First, a large set of CFD simulations are conducted using the COMSOL Multiphysics computer program for parameters selected on a three (or six) dimensional Latin hypercube. The calculations store the coefficients of lift and drag which are used to train the MLPNNs. After training, the networks are validated against separate unseen data sets, and good agreement is observed. The parametric variation of the lift and drag coefficients for a single cylinder are presented for different Reynolds numbers, angle of attack, and aspect ratio of the cylinder.
\par A detailed sensitivity analysis is carried out using the MLPNN for lift and drag. The data sample for the sensitivity studies using the Monte-Carlo method is generated by the neural networks. The sample size is increased exponentially until the sensitivity coefficient converges to a value independent of the sample size. It can be seen that using the neural network as a surrogate model is essential since running a hundred thousand numerical simulations is computationally expensive. The study indicates which parameters are most influential over others in impacting the output variables. For a single cylinder, the lift coefficient is strongly dependent on the angle of attack, while the drag is a strong function of the Reynolds number. For the case of tandem cylinders, the lift on the upstream cylinder is most sensitive to its angle of attack, while drag is equally sensitive to the Reynolds number and upstream angle of attack. The lift on the downstream cylinder is a strong function of the angle of attack of the upstream cylinder, while the drag on the downstream cylinder is most sensitive to the upstream cylinder aspect ratio.
\par The current study has been limited to the steady region. Efforts are underway to consider supercritical Reynolds number for which the flow will become unsteady with periodic shedding of vortices.

%\section*{Data Availability Statement}
%The data that support the findings of this study are available from the corresponding author upon reasonable request.
\bibliography{References}

\end{document}